\documentclass[aps,a4paper,superscriptaddress,twocolumn,showpacs,preprintnumbers,amsmath,amssymb]{revtex4}

\usepackage{graphicx}
\usepackage{dcolumn}
\usepackage{bm}
\usepackage{amsmath}
\usepackage{amsthm}
\usepackage{hyperref}
\usepackage{dsfont}
\usepackage{color}

\begin{document}


\title{Feedback control of coherent spin states \\ using weak nondestructive measurements}


\author{T.~Vanderbruggen}
\altaffiliation[Now at: ]{ICFO --- Institut de Ci{\`e}ncies Fot{\`o}niques,
E-08860, Castelldefels Barcelona, Spain}
\email{thomas.vanderbruggen@icfo.es}
\author{R.~Kohlhaas}
\affiliation{Laboratoire Charles Fabry, Institut d'Optique, CNRS, Universit{\'e} Paris-Sud, Campus Polytechnique, RD 128, 91127 Palaiseau cedex, France}
\author{A.~Bertoldi}
\author{E.~Cantin}
\affiliation{Laboratoire Photonique, Num{\'e}rique et Nanosciences - LP2N Universit{\'e} Bordeaux - IOGS - CNRS : UMR 5298 - rue F.~Mitterand, Talence, France}
\author{A.~Landragin}
\affiliation{LNE-SYRTE, Observatoire de Paris, CNRS and UPMC,\\ 61 avenue de l'Observatoire, F-75014 Paris, France}
\author{P.~Bouyer}
\affiliation{Laboratoire Photonique, Num{\'e}rique et Nanosciences - LP2N Universit{\'e} Bordeaux - IOGS - CNRS : UMR 5298 - B{\^a}timent A30, 351 cours de la liberation, Talence, France}

\date{\today}

\begin{abstract}
We consider the decoherence of a pseudo-spin ensemble under collective random rotations, and study, both theoretically and experimentally, how a nondestructive measurement combined with real-time feedback correction can protect the state against such a decoherence process. We theoretically characterize the feedback efficiency with different parameters --- coherence, entropy, fidelity --- and show that a maximum efficiency is reached in the weak measurement regime, when the projection of the state induced by the measurement is negligible. This article presents in detail the experimental results published in [Phys. Rev. Lett. \textbf{110}, 210503 (2013)], where the feedback scheme stabilizes coherent spin states of trapped ultra-cold atoms, and nondestructively probed with a dispersive optical detection. In addition, we study the influence of several parameters, such as atom number and rotation angle, on the performance of the method. We analyze the various decoherence sources limiting the feedback efficiency and propose how to mitigate their effect. The results demonstrate the potential of the method for the real-time coherent control of atom interferometers.
\end{abstract}



\maketitle

\section{Introduction}

Nondestructive measurements of atomic samples are finding an increasing number of applications, especially for metrological purposes since atoms are at the heart of many sensors \cite{kitching2011} such as clocks, magnetometers, gravimeters and gyrometers. Nondestructive measurements show a strongly reduced heating rate compared to fluorescence or absorption probing techniques. For this reason, they are exploited, for example, to increase the cycling rate in optical lattice clocks \cite{lodewyck2009}, thus reducing the Dick effect, or to suppress the atom number fluctuations occurring in the successive preparations of cold atomic samples \cite{sawyer2012,gajdacz2013}.

Nondestructive measurements preserving the coherence of the atomic sample are of particular interest in atom interferometric sensors, which rely on the wavepacket coherence. The coherence preserving feature has been demonstrated by the real-time observation of Rabi oscillations \cite{windpassinger2008bis,chaudhury2006,bernon2011} and used for state tomography \cite{smith2006}. Moreover, when these nondestructive measurements are sensitive to the quantum fluctuations of the coherent atomic state, they prepare spin-squeezed states \cite{appel08,takano2009,schleier-smith10,chen2011,sewell2011,PhysRevLett.110.163602,sewell2013}, and allow the operation of atomic clocks beyond the shot-noise limit \cite{louchet-chauvet2010,leroux10}.  

This publication is a companion article of Ref.~\cite{vanderbruggen2013}, where the real-time feedback control of a collective pseudo-spin is considered. As a proof of principle, we present how a feedback control based on a nondestructive measurement protects the state against the decoherence induced by random rotations of the collective spin. The feedback scheme is similar to the method proposed for a single qubit in Ref.~\cite{branczyk2007} and later implemented experimentally with photonic qubits \cite{gillett2010}. It demonstrates that weak measurements monitoring the disturbance caused by the environment can protect the coherence of collective quantum systems \cite{lloyd2000}. Our work can serve as a basis for experiments on the coherent feedback control of atomic interferometers, and provides a method to estimate the potential of such feedback systems.  

The article is organized as follows. In Sec.~\ref{sec:th_analysis}, a theoretical introduction to the problem is presented. After recalling the concept of collective pseudo-spin and the related collective unitary evolutions, we introduce the decoherence process resulting from random collective rotations (RCRs). More particularly, we present the specific case of a binary RCR, which constitute a practical benchmark used throughout this article to understand and characterize a feedback system on a simple situation. We then model the nondestructive measurement with a Gaussian measurement operator and study the feedback controller in the weak measurement limit. To characterize the feedback efficiency, we compare three methods based on the coherence, the entropy and the fidelity, respectively. Using a Monte-Carlo simulation, we study the crossover between the weak and strong measurement limits, and show that the best efficiency is reached using a weak measurement, for the considered controller. Finally, we analyze the feedback control of a more general kind of collective decoherence, called an analog RCR.

Sec.~\ref{sec:exp_impl} describes the experimental implementation of the feedback scheme with cold $^{87}$Rb atoms and a dispersive heterodyne detection. After a brief presentation of the dipole trap and of the atomic state preparation, we explain the implementation of the RCR and the feedback controller with resonant microwave pulses. It is then shown how a nondestructive probe based on the frequency modulation spectroscopy technique can measure the pseudo-spin collective observable $J_{z}$, and how it is possible to cancel the inhomogeneous light-shift of the probe \cite{windpassinger2008} and the related decoherence --- a mandatory condition to implement the control scheme. We also characterize the decoherence resulting from the inhomogeneous differential light-shift induced by the trapping beam on the clock transition, and show this is not a limitation for the experiments presented here.

Finally, in Sec.~\ref{sec:results} we present the experimental results and focus on the data acquisition and analysis. First, from the study of a binary RCR followed by a correction, we analyze the influence of the atom number and of the probe strength on the remaining coherence, using a Ramsey-like measurement. The iteration of the sequence of binary RCRs and feedbacks demonstrates how the collective state can be protected over time. To conclude the experimental realization of an analog RCR is presented.

\section{\label{sec:th_analysis}Theoretical background}

\begin{figure}[!h]
\begin{center}
\includegraphics[width=8.5cm,keepaspectratio]{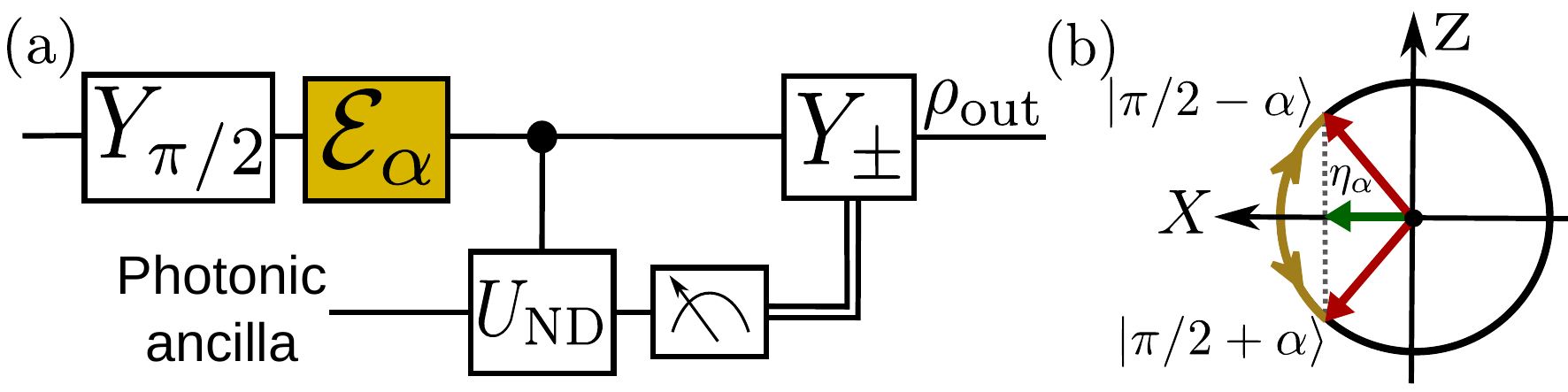}
\caption{(Color online) (a) Scheme of the feedback sequence for a binary RCR. An initial rotation $Y_{\pi/2}$ prepares a coherent superposition which later undergoes a random collective rotation $\mathcal{E}_{\alpha}$. The system is then indirectly measured using a photonic probe that evolves according to a nondestructive interaction $U_{\rm ND}$. The final rotation $Y_{\pm}$, implementing the correction, is conditioned to the measurement result. (b) Reduction of the collective spin coherence by the random binary rotation of angle $\alpha$. The coherence of the statistical mixture after the noise action is the length of the projection of the collective Bloch vector on the $X$ axis of the sphere.}
\label{fig:coherence_bit_phase_flip}
\end{center}
\end{figure}

In this part, we present the decoherence of a coherent spin state induced by the RCR noise model, and its successive recovery based on a nondestructive detection combined with feedback control [Fig.~\ref{fig:coherence_bit_phase_flip}~(a)]. To quantify the correction efficiency, different criteria are defined and compared using both analytical results obtained in the weak measurement limit, and numerical simulations. In particular, we analyze the efficiency versus the measurement backaction and show that the maximum efficiency is reached in the weak-measurement regime for all the considered criteria.

\subsection{Coherent spin-state and collective rotations}

The system considered is an ensemble of $N_{\rm at} = 2 j$ spin-1/2 indistinguishable particles. From the spin operators $( \sigma_{x}^{(i)}, \sigma_{y}^{(i)}, \sigma_{z}^{(i)} )$ related to the $i$th particle, the collective operators $( J_{x}, J_{y}, J_{z} )$ are built according to $J_{k} = \sum_{i} \sigma_{k}^{(i)}$. They are spin operators since they satisfy the commutation relations $\left[ J_{i}, J_{j} \right] = i \epsilon_{ijk} J_{k}$ ($\epsilon_{ijk}$ is the Levi-Civita tensor). The basis that codiagonalizes $\mathbf{J}^{2}=J_{x}^{2}+J_{y}^{2}+J_{z}^{2}$ and $J_{z}$ is called the Dicke basis $\left\{ \left|j,m \right\rangle, -j \leq m \leq j \right\}$:
\begin{eqnarray}
\mathbf{J}^{2} \left|j,m \right\rangle & = & j(j+1) \left|j,m \right\rangle, \\
J_{z} \left|j,m \right\rangle & = & m \left|j,m \right\rangle.
\end{eqnarray}
The collective spin operators are the generators of the unitary group of the collective state rotations. The rotations around each axis $X$, $Y$ and $Z$ of the Bloch sphere are $X_{\gamma} = e^{i \gamma J_{x}}$, $Y_{\theta} = e^{i \theta J_{y}}$ and $Z_{\varphi} = e^{i \varphi J_{z}}$, respectively. The state $\left| \theta, \varphi \right\rangle = Z_{\varphi} Y_{\theta} \left|j,-j \right\rangle$ is the coherent spin state (CSS) pointing in the $\left( \theta, \varphi \right)$ direction of the Bloch sphere. To simplify the notations, when $\varphi=0$ the CSS is written as $\left| \theta \right\rangle \equiv \left| \theta, 0 \right\rangle$.

The decoherence of a CSS can occur in two main ways: either the pointing direction of the state is disturbed, or the state leaks out of the maximal $J = N/2$ CSS. Here we present a method to correct the coherence loss caused by the first mechanism. The correction method relies on a measurement, which in turns induces a partial decoherence due to the second mechanism, and a consecutive retroaction on the spin direction. This approach is investigated both theoretically and experimentally, and we show that, for a suitable trade-off, it can improve the state coherence.


\subsection{Binary RCRs and decoherence}

The initial state is the CSS $\left| \psi_{0} \right\rangle \equiv \left| \pi/2 \right\rangle$. The state $\left| \psi_{0} \right\rangle$ experiences a \textit{binary} RCR that consists in a rotation of angle $\alpha$ with a random direction around the $Y$ axis of the Bloch sphere. The map of this process $\rho \mapsto \mathcal{E}_{\alpha} \left( \rho \right)$ is characterized by the following Kraus decomposition:
\begin{equation}
\mathcal{E}_{\alpha} \left( \rho \right) = \frac{1}{2} Y_{\alpha} \rho Y_{\alpha}^{\dagger} + \frac{1}{2} Y_{-\alpha} \rho Y_{-\alpha}^{\dagger}.
\end{equation}

The RCR creates a statistical mixture of the states $\left| \pi/2 + \alpha \right\rangle$ and $\left| \pi/2 - \alpha \right\rangle$, as depicted in Fig.~\ref{fig:coherence_bit_phase_flip}~(b). More precisely, the density operator generated by the process $\mathcal{E}_{\alpha}$ is:
\begin{equation}
\mathcal{E}_{\alpha} \left( \rho_{0} \right) = \frac{1}{2} \rho \left( \frac{\pi}{2} + \alpha \right) + \frac{1}{2} \rho \left( \frac{\pi}{2} - \alpha \right),
\label{eq:rho_bit_flip}
\end{equation}
where $\rho_{0} \equiv \left| \psi_{0} \right\rangle \left\langle \psi_{0} \right|$ and $\rho \left( \theta \right) \equiv \left| \theta \right\rangle \left\langle \theta \right|$.

\subsection{Evolution under the measurement}

We consider a situation where the sample is probed with a far off-resonance optical probe. The detection has a finite resolution $\sigma_{\rm det}$ and the evolution of the system determined by the measurement is modelled by a set of Gaussian measurement operators $\left\{ M_{m_{0}} \right\}$ of the observable $J_{z}$:
\begin{equation}
M_{m_{0}} = \left(2 \pi \sigma_{\rm det}^{2} \right)^{-1/4} \exp \left[ - \frac{1}{4 \sigma_{\rm det}^{2}} \left( J_{z} - m_{0} \right)^{2} \right].
\label{eq:meas_op}
\end{equation}
An important case is that of a shot-noise limited detection for which $\sigma_{\rm det} = M^{-1} N_{p}^{-1/2}/2$ ($M$ is the probe coupling strength depending on the coupling of the atoms to the optical probe and in particular on the on-resonance optical depth, and $N_{p}$ is the number of photons in the detection pulse \cite{vanderbruggen2011}).

The values $m_{0}$ accessible by the measurement are not bounded between $-j$ and $j$ since for a detection with low resolution the uncertainty may be arbitrary large. Moreover, $m_{0} \in \mathds{R}$ since the measurement output is a continuous parameter. In these conditions, the measurement operator satisfies the completeness relation: $\int_{-\infty}^{+\infty} M_{m_{0}} M_{m_{0}}^{\dagger} dm_{0} = \mathds{1}$, and the set $\left\{ E_{m_{0}} = M_{m_{0}} M_{m_{0}}^{\dagger} \right\}_{m_{0} \in \mathds{R}}$ is a continuous positive operator valued measurement.

If the number of atoms $N_{\rm at}$ in the sample is large then, by the Moivre-Laplace theorem (central-limit theorem for binomial distributions), a CSS $\left| \theta, \varphi \right\rangle = \sum_{m} c_{m} \left( \theta, \varphi \right) \left| j,m \right\rangle$ can be approximated with a Gaussian state
\begin{equation}
c_{m} \left( \theta, \varphi \right) = \frac{e^{-i \varphi m}}{\sqrt{\sqrt{\pi j} \sin \theta}} \exp \left[- \frac{\left( m - j \cos \theta \right)^{2}}{2 j \sin^{2} \theta} \right],
\end{equation}
and the width of this distribution is $\sigma_{\rm at} = N_{\rm at}^{1/2}/2$, that is the atomic shot-noise. When a measurement is performed on this state, the probability to obtain $m_{0}$ at the output is
\begin{align}
p \left(m_{0} | \theta , \varphi \right) & = \left\langle \theta,\varphi \right| M_{m_{0}}^{\dagger} M_{m_{0}} \left| \theta , \varphi \right\rangle \\
& = \frac{1}{\sqrt{2 \pi}} \frac{\xi_{\theta}}{\sigma_{\rm det}} \exp \left[ -\frac{\xi_{\theta}^{2} \left( m_{0} - j \cos \theta \right)^{2} }{2 \sigma_{\rm det}^{2}} \right],
\label{eq:proba_n0_css}
\end{align}
where $\xi_{\theta}^{2} = 1/ \left(1+\kappa^{2}\sin^{2} \theta\right)$ is the squeezing factor and $\kappa^{2} = \sigma_{\rm at}^{2} / \sigma_{\rm det}^{2}$ characterizes the projectivity of the measurement as the square of the ratio between the width of the atomic wavefunction and the resolution of the detection. If $\kappa^{2} \ll 1$ the projectivity is negligible and the measurement is said to be \textit{weak}. Conversely, in the strong measurement limit ($\kappa^{2} \gg 1$), the measurement operator is a projector: $M_{m_{0}} = \left| j,m_{0} \right\rangle \left\langle j,m_{0} \right|$. At the crossover between these two regimes ($\kappa^{2} \sim 1$), the state is partially projected and a spin-squeezed state is prepared \cite{appel08,schleier-smith10,chen2011,sewell2011}. The measurement operator Eq.~(\ref{eq:meas_op}) thus models a nondestructive measurement with arbitrary projectivity.

\subsection{\label{sec:th_feed_cont}Feedback controller and output state}

We consider a simple feedback controller that corrects for the disturbance induced by a binary RCR. The sign of $m_{0}$ determines the hemisphere where the Bloch vector lies and is then sufficient to know which rotation the system has undergone: $Y_{-\alpha}$ if the sign is positive, else $Y_{+\alpha}$. Once the rotation sign is determined, a rotation with same angle and opposite sign can be applied to bring the system back into the initial state $\left| \psi_{0} \right\rangle$. Assuming that there is no other decoherence source from the probe pulse, this controller is modelled by the following operator sum decomposition:

\begin{eqnarray}
\mathcal{C}_{\alpha} \left( \rho \right) &=& \frac{1}{2} \int_{\mathds{R}_{-}} \Big[
p\left( m | +\alpha  \right) Y_{-\alpha} M_{m} Y_{+\alpha} \rho Y_{+\alpha}^{\dagger} M_{m}^{\dagger} Y_{-\alpha}^{\dagger} \nonumber \\
& & \;\;\;\; + p\left( m | -\alpha  \right) Y_{-\alpha} M_{m} Y_{-\alpha} \rho Y_{-\alpha}^{\dagger} M_{m}^{\dagger} Y_{-\alpha}^{\dagger}  \Big] dm \nonumber \\
& & + \frac{1}{2} \int_{\mathds{R}_{+}} \Big[  
p\left( m| -\alpha  \right) Y_{+\alpha} M_{m} Y_{-\alpha} \rho Y_{-\alpha}^{\dagger} M_{m}^{\dagger} Y_{+\alpha}^{\dagger}  \nonumber \\
& & \;\;\;\; + p\left( m| +\alpha  \right) Y_{+\alpha} M_{m} Y_{+\alpha} \rho Y_{+\alpha}^{\dagger} M_{m}^{\dagger} Y_{+\alpha}^{\dagger} \Big] dm , \nonumber \\ 
\label{eq:cont_map_bin}
\end{eqnarray}
where $p\left( m | \pm \alpha  \right)$ is the probability to measure $m$ given that the state has undergone the collective rotation $Y_{\pm \alpha}$. 

\subsubsection{Success probability}

The success probability is defined as the probability to detect a positive rotation sign given that the state experienced a rotation $Y_{-\alpha}$ and vice-versa. From Eq.~(\ref{eq:proba_n0_css}), we find:
\begin{eqnarray}
p_{s} & = & \int_{0}^{\infty} p \left(m_{0} | - \alpha \right) d m_{0} \\
& = & \frac{1}{2} \left[ 1 + \mathrm{erf} \left( \sqrt{j \xi_{\pi/2-\alpha}^{2} \kappa^{2}} \sin \alpha \right) \right]. 
\label{eq:p_success}
\end{eqnarray}
As we will see, this quantity is the key parameter to describe the output state and thus the behavior of the feedback system in the weak measurement limit.

\subsubsection{Weak measurement limit}

In the weak measurement limit ($\kappa^{2} \ll 1$), the projectivity is negligible and the measurement does not modify the state ($M_{m} \sim \mathds{1}$). As a consequence, the controller map Eq.~(\ref{eq:cont_map_bin}) becomes:
\begin{equation}
\mathcal{C}_{\alpha} \left( \rho \right) \sim p_{s} \rho + \frac{1-p_{s}}{2} \left[ Y_{2\alpha} \rho Y_{2\alpha}^{\dagger} + Y_{-2\alpha} \rho Y_{-2\alpha}^{\dagger} \right].
\label{eq:cont_map}
\end{equation}

Therefore, if the input state is the coherent superposition $\left| \pi/2 \right\rangle$, then Eq.~(\ref{eq:cont_map}) means that either the controller took the right decision with probability $p_{s}$ and the output state is $\left| \pi/2 \right\rangle$, or the decision was wrong and the output state is $\left| \pi/2 \pm 2 \alpha \right\rangle$ depending on the initial rotation induced by the RCR. The density matrix obtained after correction is $\rho_{\rm out} = \mathcal{C}_{\alpha} \left( \rho_{0} \right)$, explicitly:
\begin{equation}
\rho_{\rm out} = p_{s} \rho \left( \frac{\pi}{2} \right) + \frac{1 - p_{s}}{2} \left[ \rho \left( \frac{\pi}{2}+ 2\alpha \right) + \rho \left( \frac{\pi}{2}-2\alpha \right)\right].
\label{eq:rho_out}
\end{equation}
This state is fully determined from the knowledge of the success probability. For a perfect measurement with $p_{s}=1$, the output state is $\rho_{\rm out} = \rho_{0}$ which is the pure initial state: the feedback control thus perfectly corrects the disturbance induced by the RCR. To evaluate the controller in the case of an imperfect detection ($p_{s}<1$), we quantify its efficiency in protecting the initial state.
 
\subsection{\label{sec:feed_eff_crit}Evaluation of the feedback efficiency}

After the analysis of the retroaction process consisting in a RCR, a measurement and a feedback correction, we quantify the efficiency of the correction in terms of three different parameters characterizing the state:
\begin{description}
\item[Coherence] \textit{The correction shall increase the coherence}.\\
The coherence is defined as the norm of the mean Bloch vector normalized to the sphere radius: $\eta (\rho) \equiv \left\| \left\langle \mathbf{J} \right\rangle \right\| / j$, where $\left\langle \mathbf{J} \right\rangle = \left( \left\langle J_{x} \right\rangle, \left\langle J_{y} \right\rangle, \left\langle J_{z} \right\rangle \right)$ and $\left\langle J_{k} \right\rangle = \mathrm{Tr} \left( J_{k} \rho \right)$,

\item[Entropy] \textit{The correction shall reduce the entropy}.\\
The von Neumann entropy of a system with density operator $\rho$ is $S ( \rho ) \equiv - \mathrm{Tr} \left( \rho \log_{2} \rho \right)$,

\item[Fidelity] \textit{The correction shall increase the fidelity}.\\
The fidelity is defined as the projection of the output state $\rho_{\rm out}$ on the input one $\left| \psi_{0} \right\rangle$: $\mathcal{F} \left(\rho_{\rm out}, \left| \psi_{0} \right\rangle \right) \equiv \left\langle \psi_{0} \right| \rho_{\rm out} \left| \psi_{0} \right\rangle$.
\end{description}
Having several parameters to measure the efficiency is useful since, depending on the experimental context, one may be easier to estimate than another. But before raising experimental considerations, we study whether those three parameters lead to equivalent definitions of the efficiency.

From the results previously obtained in the weak measurement limit, we calculate the values for the three parameters at the different stages of the retroaction process: when the ensemble is in the initial state ($\left| \psi_{0} \right\rangle$), after the RCR ($\mathcal{E}_{\alpha} \left( \rho_{0} \right)$), and after the correction ($\mathcal{C}_{\alpha} \left( \rho_{0} \right)$). The results are presented in Tab.~\ref{tab:evol_grand_caract} and their derivation is detailed in App.~\ref{app:crit_param}.

\squeezetable
\begin{table}[h!]
		\begin{ruledtabular}
		\begin{tabular}{cccc}
		          & $\left| \psi_{0} \right\rangle$ & $\mathcal{E}_{\alpha} \left( \rho_{0} \right)$ & $\mathcal{C}_{\alpha} \left( \rho_{0} \right)$ \\
		\hline
	  &&&\\
		Coherence & 1                               & $\left| \cos \alpha \right|$ & $p_{s} + \left( 1-p_{s} \right) \cos 2 \alpha$  \\
		Entropy  & 0                               & 1 & $-p_{s} \log_{2} p_{s} + \left( 1 - p_{s} \right) \left[ 1 - \log_{2} \left( 1 - p_{s} \right) \right]$ \\
		Fidelity  & 1                               &  $e^{-2j \alpha^{2}}$ & $p_{s} + \left( 1 - p_{s} \right)e^{-2j \alpha^{2}}$ \\
		\end{tabular}
		\end{ruledtabular}
		\caption{Characteristic parameters quantifying the state at the beginning ($\left| \psi_{0} \right\rangle$), after the binary RCR ($\mathcal{E}_{\alpha} \left( \rho_{0} \right)$) and after the correction ($\rho_{\rm out} = \mathcal{C}_{\alpha} \left( \rho_{0} \right)$). These results are obtained in the limit of a weak measurement assuming that the binary RCR is the only decoherence source.}
		\label{tab:evol_grand_caract}
\end{table}

\begin{figure}[!h]
\begin{center}
\includegraphics[width=8.5cm,keepaspectratio]{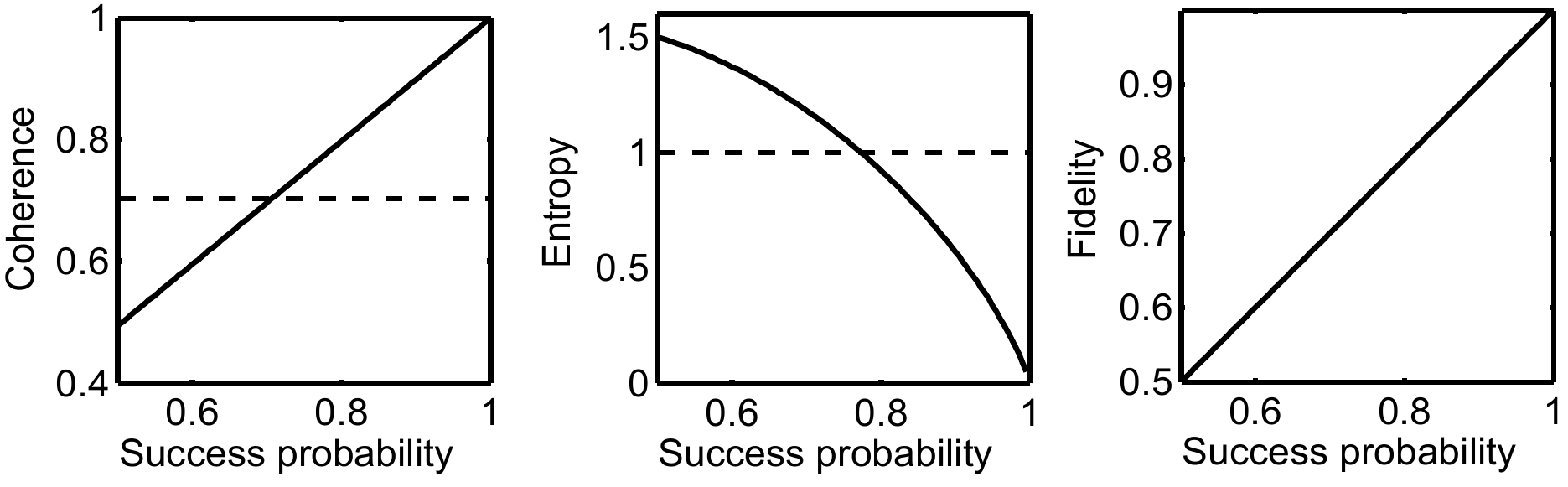}
\caption{Coherence, von Neumann entropy and fidelity of the output state $\rho_{\rm out}$ versus the success probability, for $\alpha = \pi/4$ and in the limit of a large atom number ($N_{\rm at} \gg 1$). The dotted lines indicate the reference values for the mixed state $\mathcal{E}_{\alpha} \left( \rho_{0} \right)$ generated by the RCR (for the fidelity the dotted line is not visible since it is very close to zero if the binary RCR angle satisfies $\alpha \gg N_{\rm at}^{-1/2}$) and represent the thresholds above which the correction procedure is efficient.}
\label{fig:coh_ent_fid_vs_ps}
\end{center}
\end{figure}

We already saw above that a perfect detection ($p_{s}=1$) would recover a pure state. Conversely, when the detection does not distinguish between the states $\left| \pi/2+\alpha \right\rangle$ and $\left| \pi/2-\alpha \right\rangle$ then $p_{s}=1/2$. In that case, the coherence of the output state reduces to $\eta_{\alpha}^{\rm out} = \eta_{\alpha}^{2}<\eta_{\alpha}$ and the entropy increases to $S \left( \rho_{\rm out} \right) = 3/2 > 1$. As a consequence, when the detection resolution is low the feedback action deteriorates the state instead of protecting it. We show in Fig.~\ref{fig:coh_ent_fid_vs_ps} the evolution of the coherence, the entropy and the fidelity versus the success probability for a binary RCR with angle $\alpha = \pi/4$.

We now compare the different efficiency measures by introducing the critical success probability $\tilde{p}_{s}$ above which the correction improves the relative parameter. For the fidelity, the critical probability is $\tilde{p}_{s}^{(f)} \approx 0$ (for $\alpha \gg N_{\rm at}^{-1/2}$) and the correction is always efficient; therefore the fidelity should be used with care when dealing with a large particle number system described by a Gaussian state. Concerning the coherence, the critical probability depends on the value of the RCR angle $\alpha$, more precisely for $-\pi/2 \leq \alpha \leq \pi/2$, $\tilde{p}_{s}^{(c)} ( \alpha ) = (\cos \alpha - \cos 2 \alpha)/(2 \sin^{2} \alpha)$. For $\alpha = \pi/2$, $\tilde{p}_{s}^{(c)} = 1/2$, whereas it increases to $\tilde{p}_{s}^{(c)} = 3/4$ when $\alpha = 0$. Finally, the critical probability related to the entropy is $\tilde{p}_{s}^{(e)} \sim 0.77$; it is thus more constraining than the coherence related one, even for $\alpha = 0$. The comparison of the different measures in the weak measurement regime is given by the following strict inequalities: $\tilde{p}_{s}^{(f)} < \tilde{p}_{s}^{(c)} ( \alpha ) < \tilde{p}_{s}^{(e)}$, therefore they are not equivalent. However, we will now see, using numerical simulations, that they present a similar behavior.
 
\subsection{Numerical simulations}

To study the feedback control of a binary RCR in an arbitrary measurement regime, we numerically simulate the feedback process using a Monte-Carlo analysis which generates random trajectories followed by a quantum state during the sequence RCR-measurement-correction. This allows us to verify the analytical results previously obtained in the weak measurement limit and to study the crossover between the weak and the strong measurement regime. In this simulation, the nondemolition measurement is supposed to be perfect in the sense that no decoherence (e.g. spontaneous emission or inhomogeneous light-shift) is induced by the probe beam.

\begin{figure}[!h]
\begin{center}
\includegraphics[width=8.5cm,keepaspectratio]{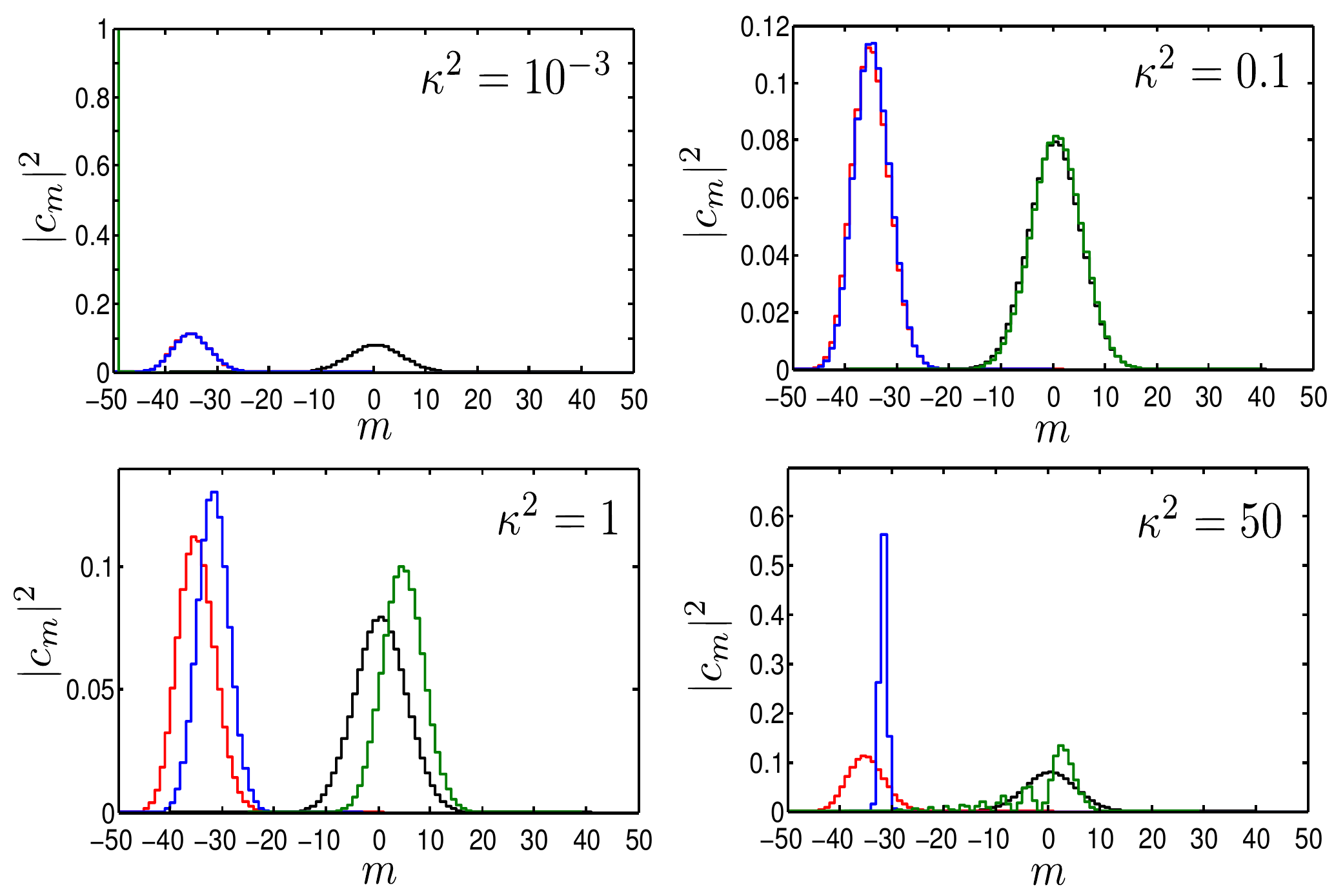}
\caption{(Color online) Distributions of the atomic wavefunction projected in the $\{ | m \rangle \}$ basis of the eigenstates of $J_{z}$ and obtained for different values of $\kappa^{2}$. The RCR angle is $\alpha = \pi / 4$ and $N_{\rm at} = 100$. The distribution are given for each step of a Monte-Carlo trajectory simulation: the initial CSS (black), after the RCR (red, light gray left), after the measurement (blue, dark gray left) and after the correction (green, light gray right). Note that in the case $\kappa^{2} = 10^{-3}$, the controller took the wrong decision and the output state is a delta peak centered in $m=-j$, meaning that the collective spin is pointing towards the south pole of the Bloch sphere.}
\label{fig:distribs_simuls}
\end{center}
\end{figure}

The simulation uses as the initial state the CSS $\left| \psi_{0} \right\rangle = \left| \pi/2 \right\rangle$, and applies a binary RCR to it, that is a rotation $Y_{\alpha}$ or $Y_{-\alpha}$ with probability $1/2$. The rotations of the collective spin are implemented using the Wigner D-matrix \cite{rose95}. From the measurement operator Eq.~(\ref{eq:meas_op}), we compute the probability density to measure $m_{0}$ and draw a value for $m_{0}$ according to this distribution. The measurement operator $M_{m_{0}}$ is then applied to compute the measurement backaction on the state. Then, depending on the sign of $m_{0}$, we apply the correction rotation. Examples of distributions obtained along a trajectory for different measurement strengths are presented in Fig.~\ref{fig:distribs_simuls}. We see that, due to the projection induced by the measurement, the distribution after the correction rotation is not Gaussian for $\kappa^{2} > 1$. 

The sequence, repeated several times starting with the same initial state, provides a statistical estimate of the success probability. The fidelity is obtained by projecting the final state on the initial one. The coherence is calculated from the average over all the simulated trajectories of the norm of the output Bloch vector. Finally, the von Neumann entropy is $S = - \mathrm{Tr} ( \overline{\rho_{\rm out}} \log_{2} \overline{\rho_{\rm out}})$ with
\begin{equation}
\overline{\rho_{\rm out}} = \frac{1}{N_{\rm traj}} \sum_{k=1}^{N_{\rm traj}} \left| \psi_{\rm out}(k) \right\rangle \left\langle \psi_{\rm out}(k) \right|,
\end{equation}
where $N_{\rm traj}$ is the number of simulated trajectories and $\left| \psi_{\rm out} (k) \right\rangle$ is the state obtained at the end of the $k$th trajectory.

\begin{figure}[!h]
\begin{center}
\includegraphics[width=8.5cm,keepaspectratio]{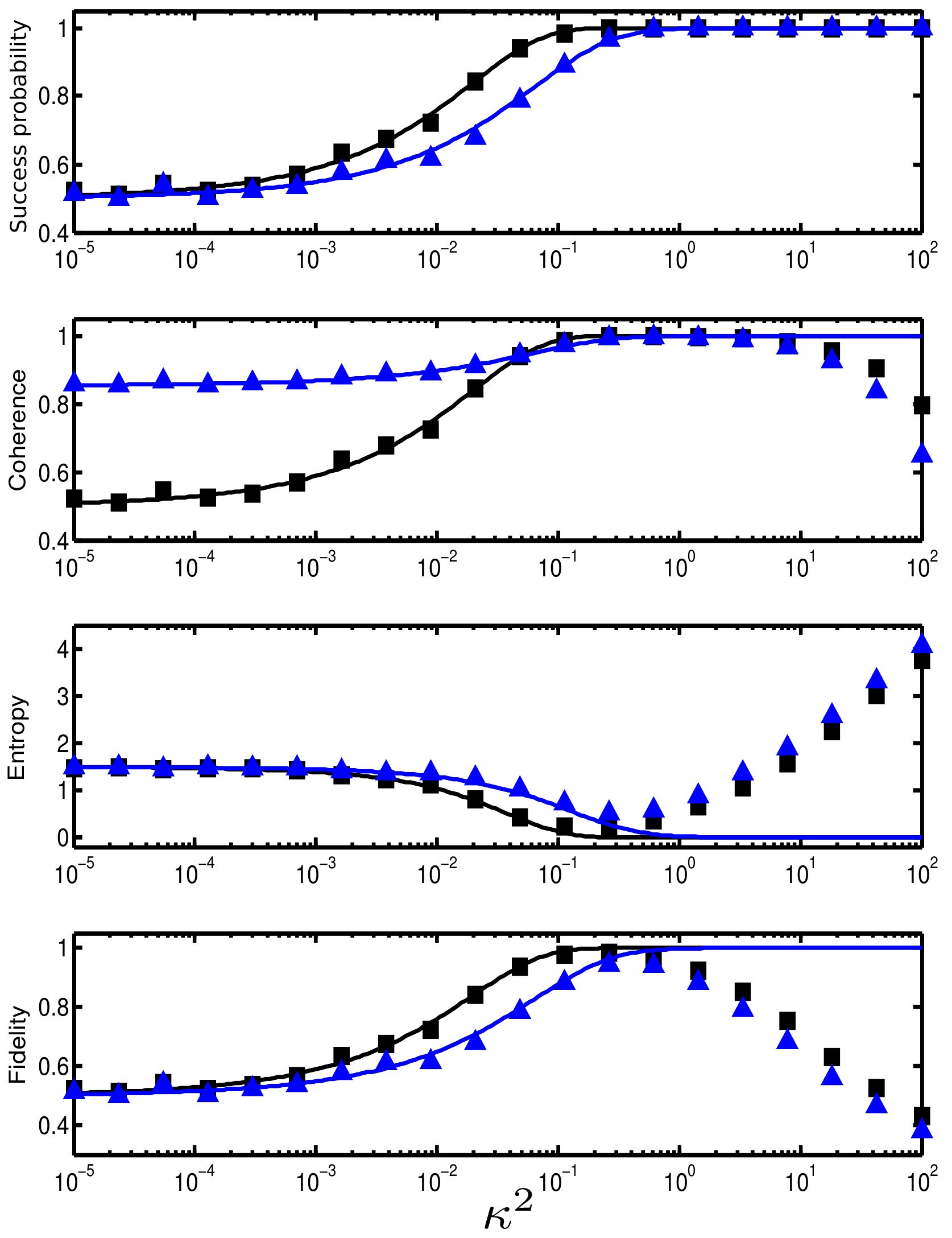}
\caption{(Color online) Results of the Monte-Carlo simulation for 1000 trajectories, where the parameters are $N_{\rm at} = 100$ and $\alpha = \pi/4$ (black squares) and $\alpha = \pi/8$ (blue triangles). From top to bottom: the success probability, the coherence, the entropy and the fidelity are plotted versus $\kappa^{2}$. The solid lines are the analytical results obtained in the weak measurement limit ($\kappa^{2} \ll 1$).}
\label{fig:MonteCarlo_N1000}
\end{center}
\end{figure} 

We run the simulation for a sample containing $N_{\rm at} = 100$ atoms and for different values of $\kappa^{2}$ spanning from a weak to a strong measurement. The results are presented in Fig.~\ref{fig:MonteCarlo_N1000}. The analytical results obtained in the weak measurement limit ($\kappa^{2} \lesssim 0.1$) are in good agreement with the simulations. That is not the case for the strong measurement regime where the analytical expressions are inappropriate because they do not consider the projection of the state resulting from the measurement. The simulation correctly predicts the increase of the entropy and the decrease of coherence and fidelity with the increasing measurement strength. We see that, even if the efficiency measures are not equivalent, an optimum is reached for a similar value of $\kappa^{2}$ for all the three parameters. Moreover, this optimum is reached for $\kappa^{2} < 1$, that is for a weak measurement.

It is useful to note that, because of the state projection, it is not possible to fully recover the target state with the chosen controller, and this even in the case of a noise resulting from unitary Kraus operators and measured with a perfectly nondestructive detection. However, the recovered state can be arbitrarily close to the target state given that the on-resonance optical density of the sample is sufficiently large.

To conclude, fidelity, entropy and coherence may all be used to evaluate the efficiency of a feedback system for collective spin states. However, as shown in Sec.~\ref{sec:feed_eff_crit}, the  fidelity is not strongly discriminant to characterize feedback schemes involving CSSs due to their quasi-orthogonality. The coherence is the parameter adopted in the experimental part of this article, since it can be directly determined from the fringe contrast at the output of a Ramsey interferometer. In the following sections, we thus consider only the coherence. 

\subsection{\label{sec:th_analog_rcr}Analog RCR}

The RCR decoherence model is now generalized to a rotation not only with random direction, but also with random angle. We call such a decoherence model an \textit{analog} RCR. For this model, the choice of the correction strategy is not trivial: in the following we analyse two possible approaches and compare them considering their effect on the coherence.

\subsubsection{Decoherence}

As a case study, we consider a RCR angle $\alpha$ uniformly distributed in $\left[-\pi/2,+\pi/2 \right]$. The continuous Kraus operators related to this analog RCR are $E_{\alpha} = Y_{\alpha}/\sqrt{\pi}$, for $-\pi/2 \leq \alpha \leq +\pi/2$, generating the map:
\begin{equation}
\mathcal{E} (\rho) = \frac{1}{\pi} \int_{-\pi/2}^{+\pi/2} Y_{\alpha} \rho Y_{\alpha}^{\dagger} \; d \alpha,
\end{equation}
and satisfies the completeness relation: $\int_{-\pi/2}^{+\pi/2} E_{\alpha} E_{\alpha}^{\dagger} = \mathds{1}$. Therefore, this decoherence process transforms the initial state $\left| \pi/2 \right\rangle$ into $\mathcal{E} (| \pi/2 \rangle) =\int_{-\pi/2}^{+\pi/2} \rho ( \pi/2 + \alpha ) d \alpha / \pi$, and the coherence of this statistical mixture is (App.~\ref{app:coh_mixt_css}):
\begin{equation}
\eta \left[ \mathcal{E} (\left| \pi/2 \right\rangle) \right] = \frac{1}{\pi} \left| \int_{-\pi/2}^{+\pi/2} e^{i \alpha} \, d\alpha \right| = \frac{2}{\pi}.
\end{equation}
The coherence is reduced to about 63~\% because of the specific analog RCR considered.

\subsubsection{Controller}

The controller is described by a map that sums over all the possible RCR angles $\alpha$ and possible measurement outcomes in a sequence noise-measurement-correction:
\begin{equation}
\mathcal{C} (\rho) = \int \limits_{-\pi/2}^{+\pi/2} \frac{d\alpha}{\pi} \int \limits_{-\infty}^{+\infty} dz \; p \left( z | \alpha \right) Y_{\Theta_{g}(z)} M_{z} Y_{\alpha} \rho Y_{\alpha}^{\dagger} M_{z}^{\dagger} Y_{\Theta_{g}(z)}^{\dagger},
\end{equation}
where $z \equiv m_{0}/j$ is the measurement output normalized to the Bloch sphere radius, and $\Theta_{g}(z)$ is the correction angle depending on the measurement result $z$. The conditional probability to measure $z$ given that the state has been rotated of an angle $\alpha$ around the $Y$ axis is obtained from Eq.~(\ref{eq:proba_n0_css}):
\begin{equation}
p \left( z | \alpha \right) = \frac{1}{\sqrt{2 \pi \sigma_{\alpha}^{2}}} \exp \left[-\frac{\left( z - \sin \alpha \right)^{2}}{2 \sigma_{\alpha}^{2}}\right],
\end{equation}
where $\sigma_{\alpha} = \sigma_{\rm det} / \xi_{\pi/2-\alpha}$. For a weak measurement we can have $|z|>1$, hence we adopt the following strategy to define the correction angle:
\begin{equation}
\Theta_{g}(z) = \left\{
\begin{array}{ccc}
-g \arcsin z & \mathrm{for} & |z| \leq 1 \\
-g \pi/2 & \mathrm{for} & z > 1 \\
+g \pi/2 & \mathrm{for} & z < 1
\end{array}
\right.,
\end{equation}
where $g$ is the feedback gain.

\subsubsection{Output state in the weak measurement limit}

In the weak measurement limit ($\kappa^{2} \ll 1$), $M_{z} \sim \mathds{1}$ and the measurement resolution $\sigma_{\alpha}\ \sim \sigma_{\rm det}$ is independent of $\alpha$. If the input state is the coherent superposition $\left| \pi/2 \right\rangle$, then the output state is:
\begin{equation}
\rho_{\rm out} = \frac{1}{\pi} \int \limits_{-\pi/2}^{+\pi/2} d\alpha \int \limits_{-\infty}^{+\infty} dz \; p(z|\alpha) \rho \left( \frac{\pi}{2} + \Theta_{g}(z) + \alpha \right).
\end{equation}
The coherence of this state can be written as (App.~\ref{app:coh_mixt_css}):
\begin{equation}
\eta_{\rm out} = \frac{1}{\pi} \left| \int_{-\pi/2}^{+\pi/2} d\alpha \int_{-\infty}^{+\infty} dz \; p(z|\alpha) e^{ i \left( \alpha + \Theta_{g} (z) \right) } \right|.
\end{equation}

\begin{figure}[!h]
\begin{center}
\includegraphics[width=8cm,keepaspectratio]{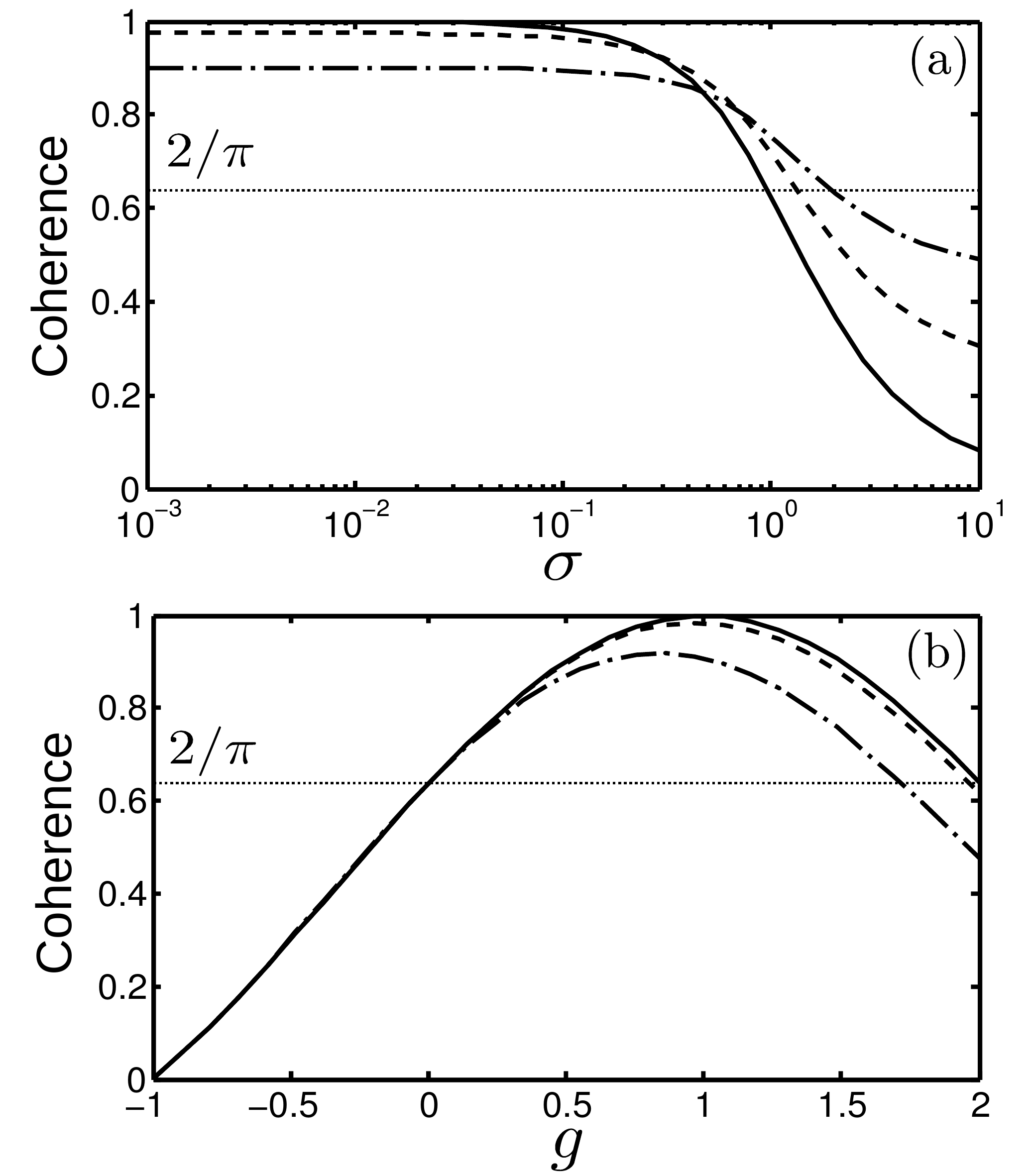}
\caption{(Color online) (a) Coherence of the output state versus the measurement resolution $\sigma$ for $g=1$ (solid line), $g=3/4$ (dashed line) and $g=1/2$ (dot-dashed line). (b) Coherence versus the feedback gain $g$ for $\sigma = 0$ (solid line), $\sigma = 1/10$ (dashed line) and $\sigma=1/3$ (dot-dashed line). For both graphs, the dotted line is the remaining coherence after the analog RCR.}
\label{fig:coherence_analog_feedback}
\end{center}
\end{figure}

We depicted Fig.~\ref{fig:coherence_analog_feedback}(a) the variation of the coherence versus the measurement resolution for different feedback gains. We see that at high resolution, a low gain does not allow a full coherence recovery. However, in a situation where the resolution is low, a reduced gain provides better performances. This reflects in Fig.~\ref{fig:coherence_analog_feedback}(b): the optimum gain is below unity for a detection with finite resolution. It results from a compromise between the strength of the correction and the effect of the detection noise mapped onto the final state through the feedback process. For example, if $\sigma=1/3$ a maximum output coherence of about $0.85$ is reached for $g \sim 0.75$.

\subsubsection{\label{sec:corr_strat_2}Alternative correction strategy}

We consider an alternative strategy adopting a correction angle proportional to the measurement output: $\Theta_{g}(z) = -gz$, which avoids the saturation problem for $|z|>1$. Moreover, it offers a simplified experimental implementation of the controller since it consists only in a proportional gain.

\begin{figure}[!h]
\begin{center}
\includegraphics[width=7cm,keepaspectratio]{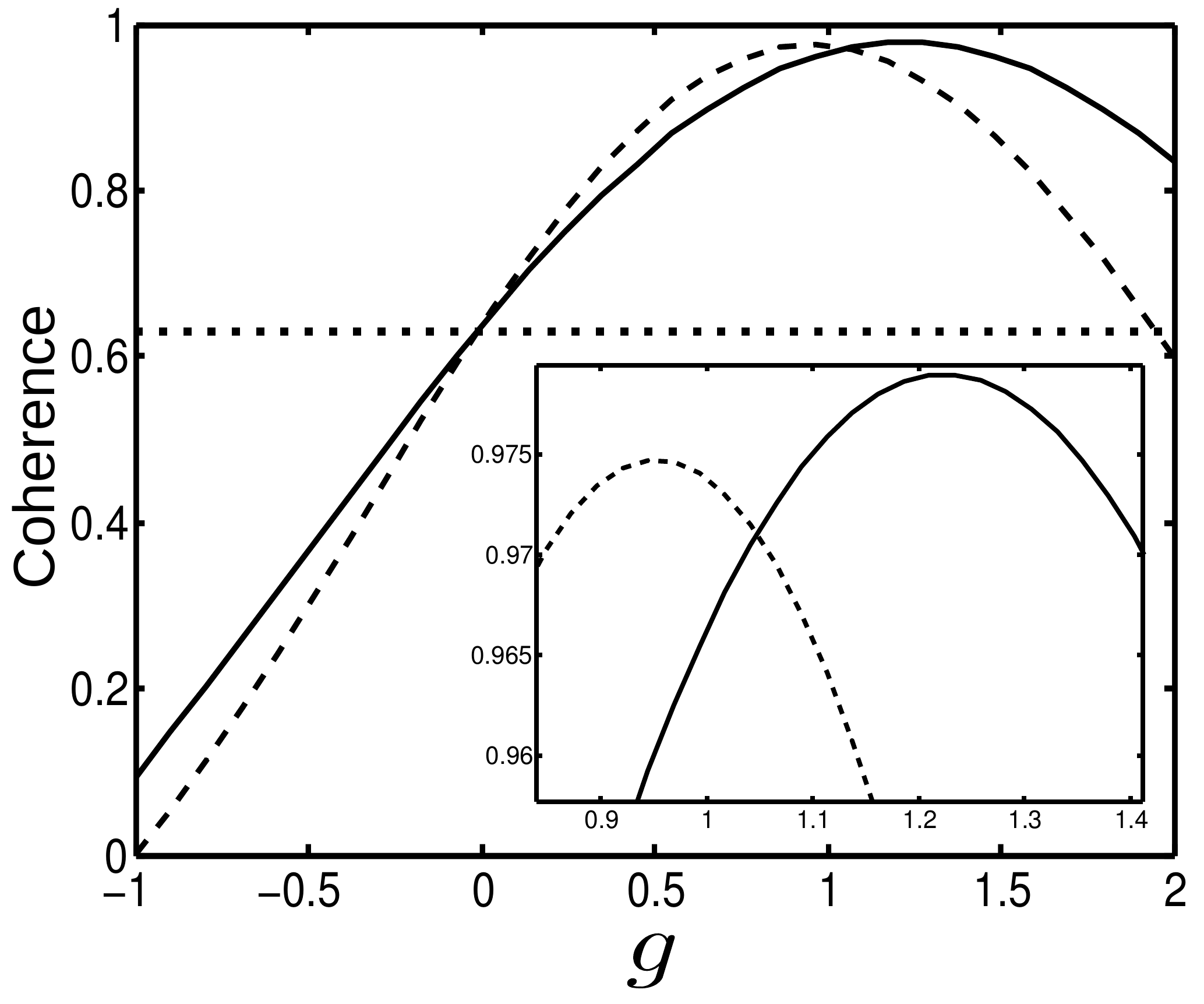}
\caption{Coherence of the output state versus the feedback gain for the two correction strategies with a detection resolution $\sigma=0.14$. The dashed black line corresponds to the case where the Bloch sphere curvature is compensated and the result is truncated for a measurement result $|z|>1$. The solid line corresponds to a correction angle proportional to the measurement result $z$. The horizontal dotted line is the remaining coherence after the analog RCR. The inset is a zoom around the optimum position.}
\label{fig:comp_strat}
\end{center}
\end{figure}

A comparison of the two strategies is presented in Fig.~\ref{fig:comp_strat} for $\sigma=0.14$, corresponding to the measurement resolution in Sec.~\ref{corr_analog_RCR}. The second strategy leads to a better result: with the first strategy a coherence of 0.975 is recovered for an optimum gain $g=0.95$ whereas it reaches 0.979 for $g=1.22$ with the second strategy.

In that case, the optimum is reached for $g \geq 1$ since $\left| z \right|  = \left| \sin \alpha \right| \leq \left| \alpha \right|$. The angle $\alpha_{0}$ for which this gain is optimized satisfies $\alpha_{0} - g z_{0} = 0$ and since $z_{0}=\sin \alpha_{0}$:
\begin{equation}
g = \frac{\alpha_{0}}{\sin \alpha_{0}}.
\end{equation}
For $g=1.22$, we find $\alpha_{0} \sim \pi/2.9$ and the gain can be experimentally adjusted by minimizing the angular spread after the correction applied to a CSS pointing in the direction $(\theta = \alpha_{0}, \varphi=0)$.

The correction method can be adapted to different kinds of RCRs. In particular, design of an optimized strategy would benefit from the prior knowledge of the angular distribution produced by a given RCR.

\section{\label{sec:exp_impl}Experimental implementation}

\begin{figure}[!h]
\begin{center}
\includegraphics[width=7cm,keepaspectratio]{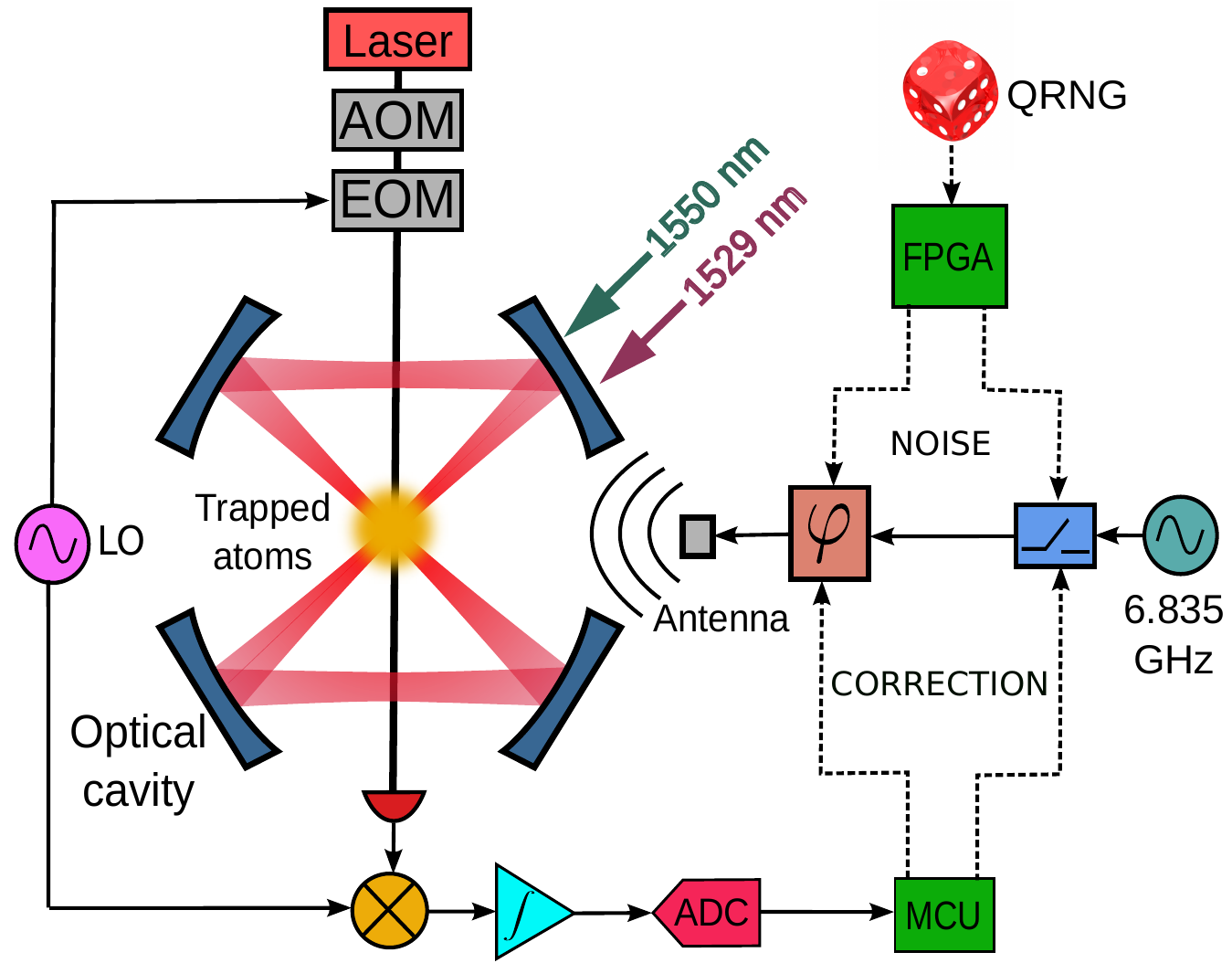}
\caption{(Color online) Experimental setup. The atomic sample is confined in a dipole trap at the crossing point of the two arms of an optical cavity in a butterfly configuration. A field programmable gate array (FPGA) controls the microwave field inducing the RCR. A micro-controller (MCU) computes the correction pulse from the nondestructive measurements. Acronyms are: quantum random number generator (QRNG), local oscillator (LO), acousto-optic modulator (AOM), electro-optic modulator (EOM) and analog-to-digital converter (ADC).}
\label{fig:setup}
\end{center}
\end{figure}

A scheme of the experimental setup implementing the feedback control is presented in Fig.~\ref{fig:setup}. A detailed description of the initial state preparation in the cavity enhanced dipole trap and of the nondestructive detection can be found in Refs.~\cite{bernon2011, vanderbruggen2013}. Here we focus on the features which are specific to the feedback application, mainly the implementation of the RCR and of the controller, and how the $J_{z}$ observable is measured with the nondestructive probe. We also analyse the main decoherence sources: we show how the inhomogeneous light-shift from the probe can be cancelled and we quantify the decoherence induced by the dipole trap radiation.

\subsection{Dipole trap and state preparation}

The atomic sample is trapped at the center of an optical cavity in a butterfly configuration \cite{bernon2011}. The cavity is injected with a radiation at 1550~nm to generate the optical trap. Due to the $5^{2} \mathrm{P}_{3/2} \rightarrow 4^{2} \mathrm{D}_{5/2,3/2}$ transitions at 1529.3~nm, the red detuned trapping beam induces a spatially inhomogeneous light-shift on the D$_{2}$ transition used for the nondestructive probing \cite{bertoldi2010}. Since a precise frequency setting of the nondestructive probe detunings is required, this light-shift is compensated by injecting the cavity, using the serrodyne modulation technique \cite{kohlhaas2012}, with a radiation at 1528.7~nm blue detuned with the $5^{2} \mathrm{P}_{3/2} \rightarrow 4^{2} \mathrm{D}_{5/2,3/2}$ transitions.

After the loading of the atomic ensemble in the dipole trap, the intracavity power is ramped down in 130~ms from 200~W to about 10~W per cavity arm to evaporatively cool the atomic sample down to a temperature of 10~$\mu$K. The radius at $1/e^{2}$ of the trapped cloud is 50~$\mu$m. The atoms are initially trapped in the $\left| 5^{2} \mathrm{S}_{1/2}, F=1 \right\rangle$ hyperfine state, and a bias magnetic field of 0.5~Gauss is applied in the direction parallel to the polarization of the nondestructive probe. The procedure to prepare the sample in the $\left| F=1, m_{F}=0 \right\rangle$ state begins with a microwave $\pi$ pulse followed by a light pulse on the $\left| F=1 \right\rangle \rightarrow \left| F'=2 \right\rangle$ transition to repump the residual population from the $\left| F=1 \right\rangle$ to the $\left| F=2 \right\rangle$ level; then about one third of the atoms are in the $\left| F=2, m_{F}=0 \right\rangle$ state. A second $\pi$ pulse is applied to populate only the $m_{F} = 0$ sublevel of the $\left| F=1 \right\rangle$ state. The residual atoms in the $\left| F=2 \right\rangle$ level are expelled from the trap using light tuned on the cycling transition $\left| F=2 \right\rangle \rightarrow \left| F'=3 \right\rangle$. To increase further the purity of the sample, the whole sequence is repeated twice. We characterized the prepared state using an absorption imaging technique and we measured that the cloud contains about $5 \times 10^{5}$ atoms and more than 99~\% of them are polarized in the $\left| F=1, m_{F}=0 \right\rangle$ state.

\subsection{Random rotation and controller implementation}

The pseudo-spin is realized by the two-level system with eigenstates $| 0 \rangle \equiv \left|5^{2} \mathrm{S}_{1/2},F=1,m_{F}=0 \right\rangle$ and $| 1 \rangle \equiv \left|5^{2} \mathrm{S}_{1/2},F=2,m_{F}=0 \right\rangle$. A resonant microwave radiation prepares an arbitrary coherent superposition of these two states. The microwave source is composed of a 7~GHz microwave oscillator mixed with a 166~MHz rf source to obtain the 6.834~GHz signal resonant with the pseudo-spin transition. A Rabi oscillations measurement is used to determine the $\pi$ pulse duration: $\tau_{\pi} = 151.2 \pm 0.2 \; \mu$s. A rf switch placed on the rf signal produces the microwave pulses. Moreover, a phase-shifter controls the phase of the microwave and thus the rotation axis of the Bloch vector. This phase-shifter is designed to implement the QPSK (Quadrature Phase-Shift Keying) telecommunication protocol, allowing us to precisely set the microwave phase to the values: $0$, $\pi/2$, $\pi$ and $-\pi/2$ with two control bits.

The RCR is implemented using a FPGA that generates the desire probability distribution from a quantum random number generator (QRNG -- Quantis, IDQuantique). The uniform distribution that generates the analog RCR as well as the sign of the binary RCR are provided by the QRNG output. 

The detection pulse is demodulated and integrated to obtain its average value. The output of the integrator is then digitized and treated with a micro-controller unit (MCU, ADuC814 from Analog Devices) which implements the feedback controller by computing the sign and the duration of the correction microwave pulse.

\subsection{Nondestructive probe}

The detection uses a far off-resonance optical probe \cite{appel08,schleier-smith10,koschorreck2010,kohnen2011} which is phase-shifted depending on the atomic population. The measurement of the phase-shift is performed by the frequency modulation spectroscopy technique: an optical carrier is modulated to produce sidebands, and one sideband is placed close to an atomic transition so that it undergoes a phase-shift proportional to the population in the probed level. The amplitude of the beatnote between the sideband and the carrier, detected on a photodiode, depends on the relative phase between these two frequency components, and therefore on the population of the probed atomic level.

In our setup (Fig.~\ref{fig:setup}) an extended cavity diode laser is frequency locked to an atomic reference. The beam passes through an acousto-optic modulator (AOM) generating the probe pulses before being phase modulated with an electro-optic modulator (EOM) feed by the local oscillator (LO) at 3.4213~GHz. After passing through the atomic cloud, the probe beam is detected on a fast photodiode and demodulated with the local oscillator \cite{bernon2011}.

\subsubsection{Measurement of the $J_{z}$ observable}

\begin{figure}[!ht]
\begin{center}
\includegraphics[width=8.5cm,keepaspectratio]{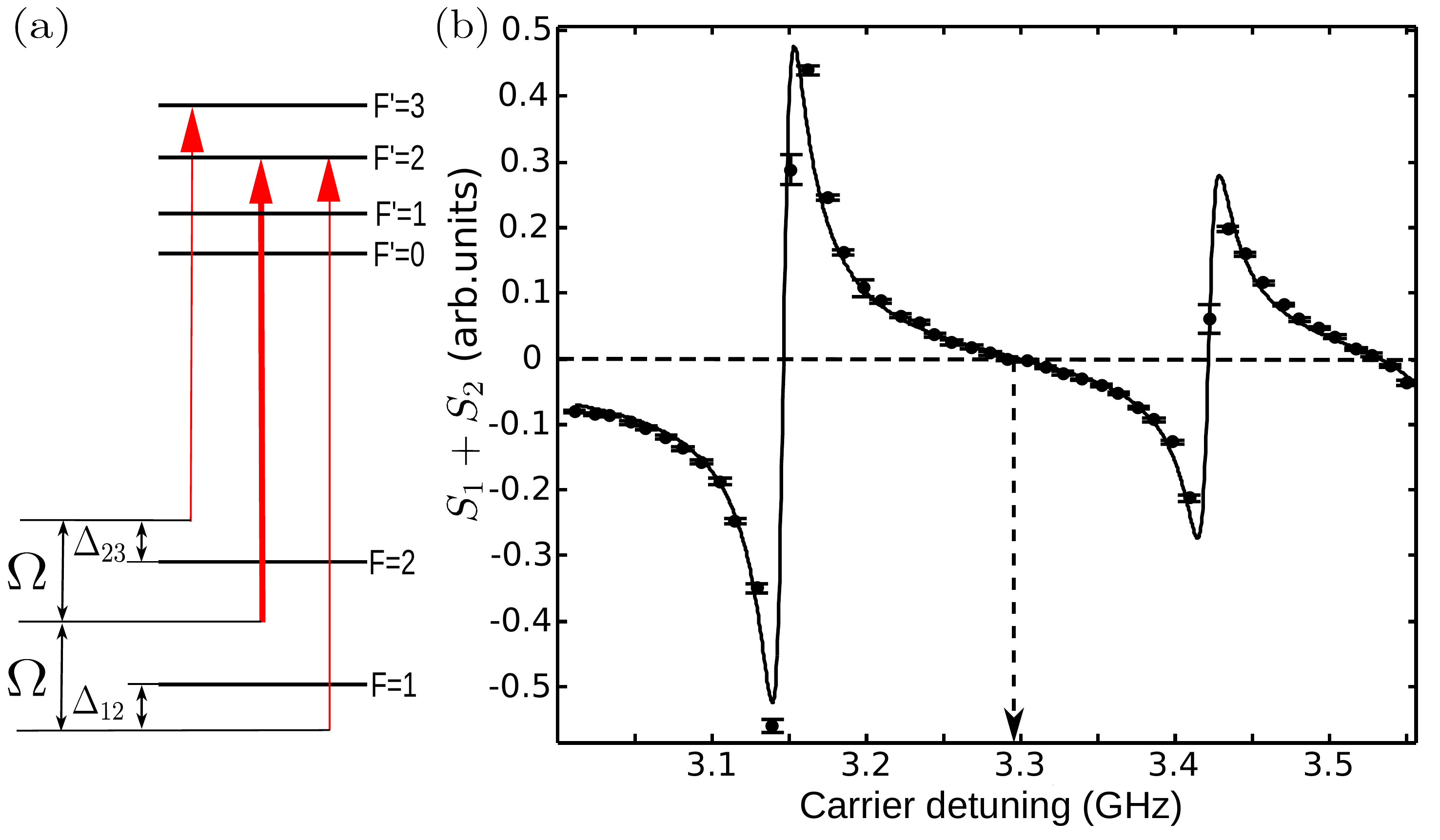}
\caption{(Color online) (a) Position of the frequency components of the probe relatively to the hyperfine structure of the $^{87}$Rb D$_{2}$ transition. The thick line is the carrier whereas the thin lines are the sidebands. (b) Coupling of the multi-frequency probe with an ensemble of $^{87}$Rb atoms versus the detuning of the carrier with respect to the $\left| F=1 \right\rangle \rightarrow \left| F'=2 \right\rangle$ transition. The modulation frequency is set to $\Omega = 3.4213$~GHz. The points with bars are the experimental results and the solid line is the coupling theoretically expected where the only adjusted parameter is a scaling factor on the amplitude.}
\label{fig:Jz_adjustement}
\end{center}
\end{figure}

With the two sidebands generated from the phase modulator it is possible to measure the collective pseudo-spin observable $J_{z}$, which is the population difference between the $\left| F=1 \right\rangle$ and the $\left| F=2 \right\rangle$ hyperfine levels: one sideband is placed close to the $\left| F=1 \right\rangle \rightarrow \left| F'=2 \right\rangle$ transition and the other one close to the $\left| F=2 \right\rangle \rightarrow \left| F'=3 \right\rangle$ transition, as depicted in Fig.~\ref{fig:Jz_adjustement}~(a). The coupling $S_{1}$ ($S_{2}$) of the first (second) sideband to the $\left| F=1 \right\rangle \rightarrow \left| F'=2 \right\rangle$ ($\left| F=2 \right\rangle \rightarrow \left| F'=3 \right\rangle$) transition satisfies
\begin{equation}
S_{F} = \sum_{F'} \frac{\gamma \Delta_{FF'}}{\Delta_{FF'}^{2} + \gamma^{2} \left( 1 + I/I_{\rm sat} \right)} S_{FF'},
\label{eq:coupling_SF}
\end{equation}
where $\gamma$ is the natural linewidth of the transition, $I$ the intensity in a single sideband, $I_{\rm sat}$ the saturation intensity of the transition and $S_{FF'}$ the dipolar coupling associated to the $\left| F \right\rangle \rightarrow \left| F' \right\rangle$ transition \cite{steck}. The phase-shift induced by the atomic sample on the probe is therefore
\begin{equation}
\phi_{\rm at} \propto N_{1} S_{1} + N_{2} S_{2},
\end{equation}
where $N_{k}$ is the population in $| F=k \rangle$. As a consequence, if the detunings $\Delta_{FF'}$ are adjusted so that $S_{1} = - S_{2}$ then $\phi_{\rm at} \propto N_{1}- N_{2}$, and the detection measures the collective observable $J_{z}$.

To establish the detunings $\Delta_{FF'}$, we first set the modulation frequency to $\Omega = 3.4213$~GHz and prepare the atoms in the coherent superposition $\left| \pi/2 \right\rangle$ so that $N_{1} = N_{2}$. We measure then the demodulated signal versus the detuning of the carrier with respect to the $\left| F=1 \right\rangle \rightarrow \left| F'=2 \right\rangle$ transition. The result in Fig.~\ref{fig:Jz_adjustement}~(b), in very good agreement with the theoretical expectation Eq.~(\ref{eq:coupling_SF}), was obtained with a carrier power of 153~$\mu$W, a power per sideband of 7.1 $\mu$W, and the beam waist of the probe at the trap position is 200~$\mu$m, which gives an intensity on the atomic sample of 11.2 mW/cm$^{2}$. Since a $\pi$ transition is probed, the saturation intensity is $I_{\rm sat} = 2.503$~mW/cm$^{2}$ \cite{steck}. The condition $S_{1} + S_{2} = 0$ is reached when the carrier is detuned by 3.291~GHz from the $\left| F=1 \right\rangle \rightarrow \left| F'=2 \right\rangle$ transition.

\subsubsection{\label{sec:meas_strength}Resolution of the detection}

\begin{figure}[!ht]
\begin{center}
\includegraphics[width=7cm,keepaspectratio]{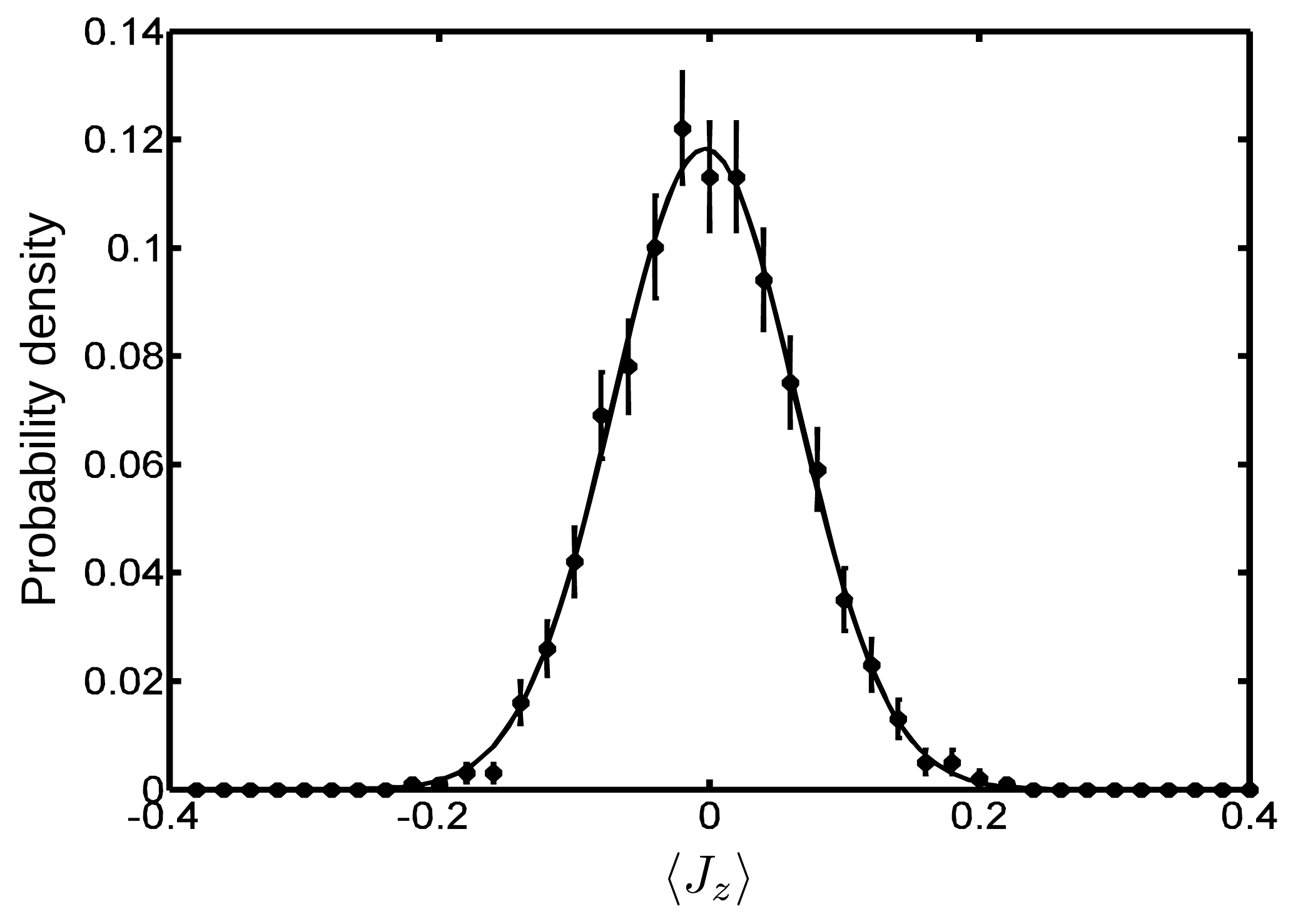}
\caption{Measurement of the detection resolution for $2.8 \times 10^{7}$ photons per sideband in a probe pulse. The points with errorbars are the experimental results obtained from 1000 measurements and the solid line is a fit with a Gaussian distribution.}
\label{fig:res_detection}
\end{center}
\end{figure}

As shown in Sec.~\ref{sec:th_feed_cont}, the measurement strength $\kappa^{2}$ strongly influences the feedback behaviour, therefore it is necessary to determine the regime in which the detection is operated. The uncertainty of the detection is obtained by performing 1000 detections of the CSS $\left| \pi/2 \right\rangle$ and the result is plotted in Fig.~\ref{fig:res_detection} for a probe pulse containing $N_{p} = 2.8 \times 10^{7}$ photons per sideband. We verify that the noise is distributed according to a Gaussian distribution with a standard deviation $\tilde{\sigma}_{\rm det} \sim 6.8 \times 10^{-2}$ in unit of the Bloch sphere radius. Since the trapped cloud contains $N_{\rm at} = 5 \times 10^{5}$ atoms after the state preparation, the noise in terms of atoms number is $\sigma_{\rm det} = \tilde{\sigma}_{\rm det} N_{\rm at} = 3.4 \times 10^{4}$. Moreover, the standard deviation of the atomic shot-noise for the CSS $\left| \pi/2 \right\rangle$ normalized to the Bloch sphere radius is $\tilde{\sigma}_{\rm at} = \sqrt{N_{\rm at}} / (2j) = 1/\sqrt{N_{\rm at}} \sim 1.4 \times 10^{-3}$. The measurement strength is thus $\kappa^{2} = \left( \tilde{\sigma}_{\rm at}/\tilde{\sigma}_{\rm det} \right)^{2} \sim 4 \times 10^{-4}$, as a consequence $\kappa^{2} \ll 1$ and the experiment is operated in the weak measurement limit.

\subsubsection{Cancellation of the probe light-shift}

The light-shift of the probe beam is often a severe limitation for the use of nondestructive methods in atom interferometry, since it induces a phase-shift that rotates the CSS around the $Z$ axis of the Bloch sphere. Moreover, a spatially inhomogeneous light-shift, arising from the intensity profile of the beam, is an additional source of decoherence. The symmetry of the frequency components in the optical probe of our detection scheme allows us to cancel the light-shift, which is a major advantage of the method.

\begin{figure}[!h]
\begin{center}
\includegraphics[width=8.5cm,keepaspectratio]{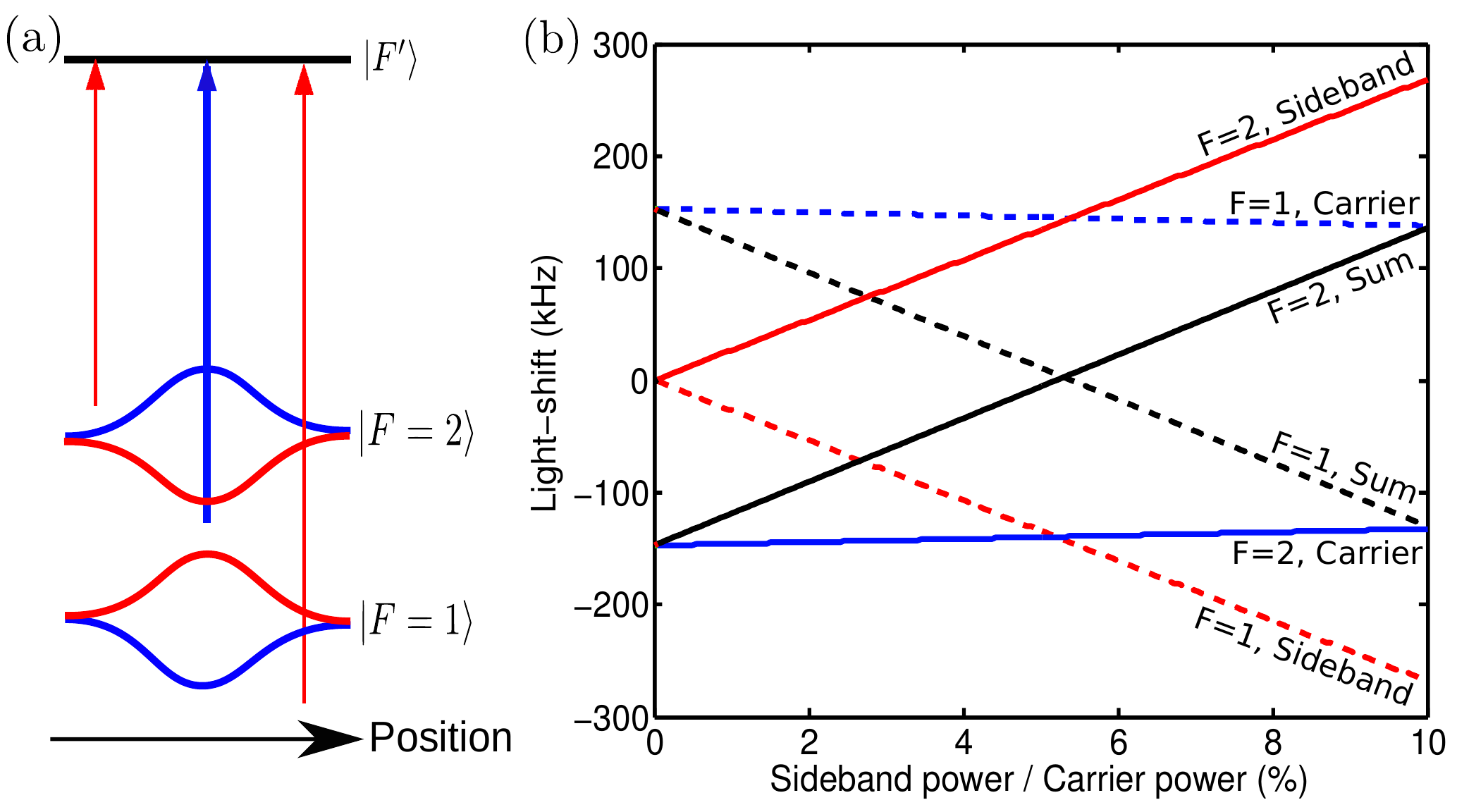}
\caption{(Color online) Compensation of the probe induced light-shift. (a) Light-shifts induced by the carrier (blue, dark gray) and by the sidebands (red, light gray). (b) Light-shifts induced by the carrier and the sidebands versus the power ratio between the sidebands and the carrier (in percent) for an overall power of 1 mW in the probe and a waist of 200 $\mu$m. The modulation frequency is $\Omega = 3.4213$ GHz and the detunings are $\Delta_{12} = 122$ MHz and $\Delta_{23} = 153$ MHz. Dashed lines are the light-shift induced on $\left| F=1 \right\rangle$ and solid lines that on $\left| F=2 \right\rangle$. The curves in blue are the light-shifts of the carrier, in red that of the sidebands and in black the resulting light-shift.}
\label{fig:light_shift_probe}
\end{center}
\end{figure}

Based on the facts that: (1) to measure the population difference the couplings of each sideband to its probed transition are the same, and (2) each sideband is on the opposite side of the transition in comparison with the carrier position; it is possible to compensate for the light-shift induced by each sideband with that induced by the carrier, as depicted in Fig.~\ref{fig:light_shift_probe}~(a). In Fig.~\ref{fig:light_shift_probe}~(b), we present a calculation of how much each frequency component contributes to the light-shift as a function of the sideband power. We observe that the compensation of the light-shift on $\left| F=1 \right\rangle$ and $\left| F=2 \right\rangle$ occurs for the same power ratio of about 5~\% between the sideband and the carrier. Moreover, since the sidebands and the carrier share the same spatial mode, the spatial compensation is perfect.

\begin{figure}[!ht]
\begin{center}
\includegraphics[width=8.5cm,keepaspectratio]{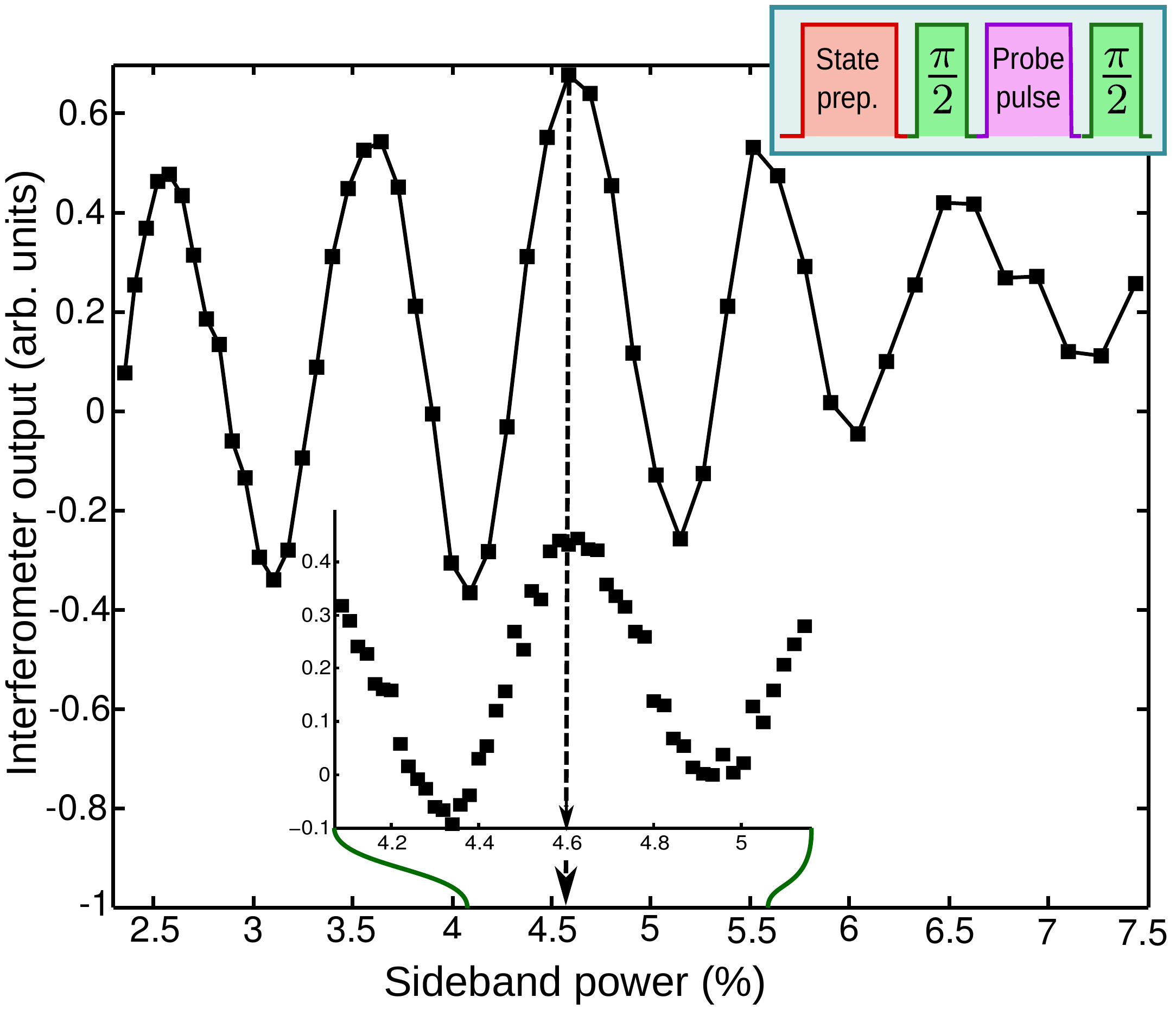}
\caption{(Color online) Output of the Ramsey interferometer versus the power in a single sideband in percent of the power in the carrier of the probe beam. A 40~$\mu$s long pulse is sent in between the two $\pi/2$ microwave pulse of the interferometer. In the lower inset, the same measurement is performed with a 70~$\mu$s long pulse providing a more precise determination of the compensation ratio. The upper inset presents the sequence used for the measurement.}
\label{fig:light_shift_comp_sidebands}
\end{center}
\end{figure}

The experimental determination of the light-shift compensation is performed with a Ramsey interferometer where a nondestructive probe pulse is sent between the two $\pi/2$ microwave pulses. If the light-shift is not compensated it induces a phase-shift on the collective spin; hence it is possible to scan interference fringes by changing the power ratio between the sidebands and the carrier, as shown in Fig.~\ref{fig:light_shift_comp_sidebands}. The observation of such Ramsey fringes provides an accurate determination and thus a precise control of the light-shift. The fringe contrast is smaller than one due to both the spontaneous emission induced by the probe and the inhomogeneous light-shift. Moreover, the fringes are not centered around $\left\langle J_{z} \right\rangle = 0$ due to the optical pumping from $\left| F=1 \right\rangle$ to $\left| F=2 \right\rangle$ that results from the spontaneous emission. The position of the zero phase-shift fringe, where the maximum contrast is achieved, provides the power required to cancel the light-shift. We determined that, for a phase modulation at 3.4213~GHz, the power ratio fulfilling the light-shift compensation is 4.6~\%, in agreement with the calculation in Fig.~\ref{fig:light_shift_probe}~(b).

\subsection{\label{sec:trap_dec}Dipole trap induced decoherence}

According to \cite{deutsch2010} and \cite{kleine2011}, the coherence evolution of a trapped spin ensemble results from two main processes: the inhomogeneous frequency shift induced by the trap profile $\Delta_{\rm T} (\mathbf{r})$, and the mean-field interaction shift $\Delta_{\rm MF} (\mathbf{r})$.

\begin{figure}[h!]
\begin{center}
\includegraphics[width=8.5cm,keepaspectratio]{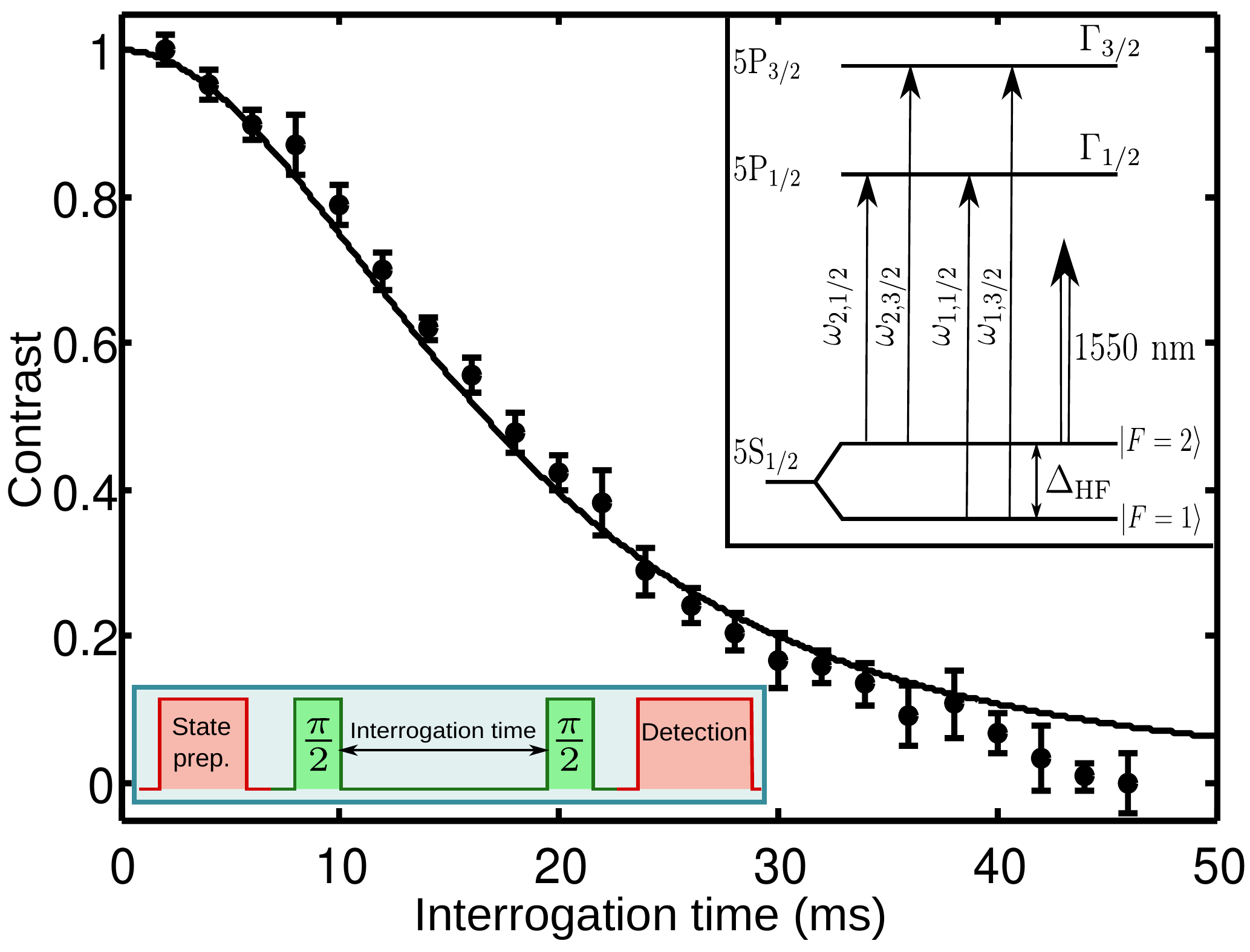}
\caption{(Color online) Dipole trap induced decoherence obtained from the contrast of the fringes at the output of a Ramsey interferometer versus the time spent by an atom in the trap. The dots are the experimental data obtained with an atomic sample at 10~$\mu$K trapped with 10~W of optical power at 1550~nm in each cavity arm. The solid line is the fit described in the text. In the upper inset the main transitions responsible for the inhomogeneous light-shift when coupled with a 1550~nm light are depicted. The lower inset shows the experimental sequence used for the measurement.}
\label{fig:contrast_vs_time}
\end{center}
\end{figure}

The dipole trap radiation at 1550~nm couples mainly to the D$_{1}$ and D$_{2}$ transitions, inducing a differential light-shift between the states $| 0 \rangle$ and $| 1 \rangle$ (see inset Fig.~\ref{fig:contrast_vs_time}). The light-shift inhomogeneity follows the Gaussian beam profile, with an amplitude of $\delta/ 2 \pi \sim 54$~Hz at the trap center (see App.~\ref{sec:trap_light_shift_clock}). As explained in \cite{deutsch2010}, the characteristic inhomogeneous shift is $\Delta_{\rm T}^{0} = \sigma_{T}^{2} \langle \partial_{x}^{2} \Delta_{\rm T} \rangle / 2$, where $\sigma_{T} = \sqrt{k_{B} T / m}/\omega$ and $\langle \partial_{x}^{2} \Delta_{\rm T} \rangle = 4 \delta / w_{0}^{2}$, with $\omega$ the trap frequency and $w_{0}$ the trap beam waist. Thus $\Delta_{\rm T}^{0} = \delta/\eta$, where $\eta = 2 k_{B} T /(m \omega^{2} w_{0}^{2})$ is the ratio between the trap potential depth and the kinetic energy of an atom. In the experimental conditions of Fig.~\ref{fig:contrast_vs_time}, $\eta \sim 4.6$ and $\Delta_{\rm T}^{0} / 2 \pi \sim 11$~Hz.

The mean-field shift satisfies $\Delta_{\rm MF}^{0} = - \gamma \bar{n}/4$, where $\bar{n}$ is the mean atomic density and $\gamma / 2 \pi = -0.48 \; \mathrm{Hz}/ 10^{12} \mathrm{cm}^{-3}$, for a clock operated between the $|0 \rangle$ and $|1 \rangle$ states of $^{87}$Rb \cite{kleine2011}. In the present situation, we have $\bar{n} \sim 4 \times 10^{12}$~cm$^{-3}$, and $\Delta_{\rm MF}^{0} / 2 \pi \sim 0.5$~Hz. Therefore $\Delta_{\rm MF}^{0} \ll \Delta_{\rm T}^{0}$, and the decoherence is dominated by the trap shift.

The evolution of the coherence is measured from the fringe contrast at the output of a Ramsey interferometer versus the trapping time $\tau$, as shown in Fig.~\ref{fig:contrast_vs_time}. Since $\Delta_{\rm MF}^{0} \ll \Delta_{\rm T}^{0}$, we follow \cite{deutsch2010} and we fit the contrast evolution with the function $\mathcal{C} (\tau) = [1 + (\Delta_{\rm T}^{0} \tau)^{2}]^{-3/2}$. We find $\Delta_{\rm T}^{0}/ 2 \pi = 7.3 \pm 0.3$~Hz, in reasonable agreement with the estimated value.

This decay time is long compared to the duration of a feedback sequence --- which is mainly set by the $\pi$ pulse duration of about 150~$\mu$s. In the present case, this decoherence source is not a limiting factor. Nevertheless, in the perspective of realizing a trapped atomic clock it may be interesting to cancel it, which is feasible using a beam at 780~nm red detuned with the $\left| F=1 \right\rangle \rightarrow \left| F' \right\rangle$ transition and blue detuned with $\left| F=2 \right\rangle \rightarrow \left| F' \right\rangle$ \cite{tackmann2011} or using the vectorial light-shift generated by an elliptical polarization \cite{dudin2010}. 

\section{\label{sec:results}Experimental results}

We now report the experimental results of the feedback control. These results expand on those published in \cite{vanderbruggen2013} in two main ways: first, we explore the behavior of the method versus more parameters, namely the atom number and the RCR angle; second, we present the data analysis of the iterated noise-measurement-correction sequence and show that the independent determination of the different decoherence sources allows us to evaluate the coherence over a large dynamic range with a reduced number of experimental cycles.

\subsection{Correction of a binary RCR}

\subsubsection{Success probability versus the atom number}

Here we analyze the influence of the number of probed atoms on the feedback efficiency by measuring how the success probability changes. In the weak measurement regime and for a binary RCR angle $\alpha=\pi/4$, the success probability [Eq.~(\ref{eq:p_success})] must satisfy
\begin{equation}
p_{s} = \frac{1}{2} \left[ 1 + \mathrm{erf} \left( \frac{N_{\rm at}}{4 \sigma_{\rm det}} \right) \right],
\label{eq:ps_vs_Nat_exp}
\end{equation}
where $\sigma_{\rm det}$ is the detection resolution. As intuition suggests, it is easier to determine the hemisphere where the spin lies when the Bloch sphere radius is large, that is for a high atom number compared to $\sigma_{\rm det}$.

\begin{figure}[!h]
\begin{center}
\includegraphics[width=8cm,keepaspectratio]{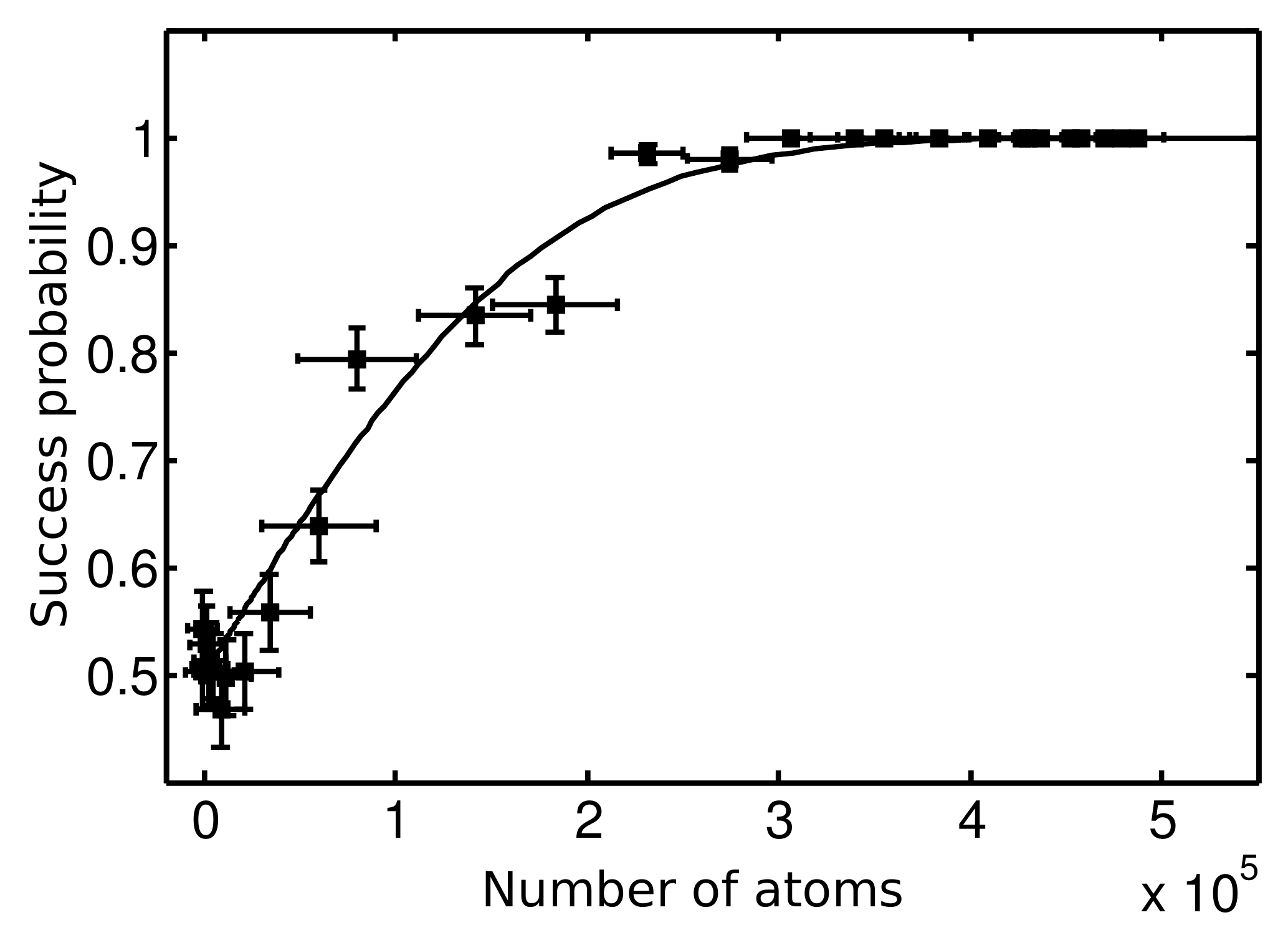}
\caption{Success probability versus the atom number in the dipole trap. The detection uses $1.4 \times 10^{7}$ photons per sideband in each pulse, the pulse duration is 1.5~$\mu$s and the RCR angle is $\alpha = \pi/4$. The dots are the experimental data and the solid line is a fit with Eq.~(\ref{eq:ps_vs_Nat_exp}).}
\label{fig:ps_vs_nat}
\end{center}
\end{figure}

We verified this behavior by measuring the success probability versus the atom number in the dipole trap. The atom number is controlled by varying the loading interval for the MOT, while the rest of the sequence remains unchanged to maintain constant the size and the temperature of the atomic cloud. The atom number in the trap is determined using absorption imaging. The success probability is measured from the repetition of 199 nondestructive detections of a binary RCR with angle $\alpha=\pi/4$. To determine the detection efficiency for a given RCR, the decisions taken by the feedback controller are compared \textit{a posteriori} with the RCR signs set by the QRNG. The result is depicted in Fig.~\ref{fig:ps_vs_nat}. The experimental data are well fitted by Eq.~(\ref{eq:ps_vs_Nat_exp}) with $\sigma_{\rm det} = 4.9 \times 10^{4}$ atoms, in agreement with the result obtained in Sec.~\ref{sec:meas_strength}.

\subsubsection{\label{sec:coh_vs_nb_photons}Coherence versus the photon number}

A compromise has to be made between the resolution of the detection, which must be high enough to detect the RCR effect, and the coherence loss due to the detection induced spontaneous emission. The trade-off is studied quantitatively by measuring the dependence of the remaining coherence after the correction from the photon number used in the detection pulse.

\begin{figure}[!h]
\begin{center}
\includegraphics[width=8.5cm,keepaspectratio]{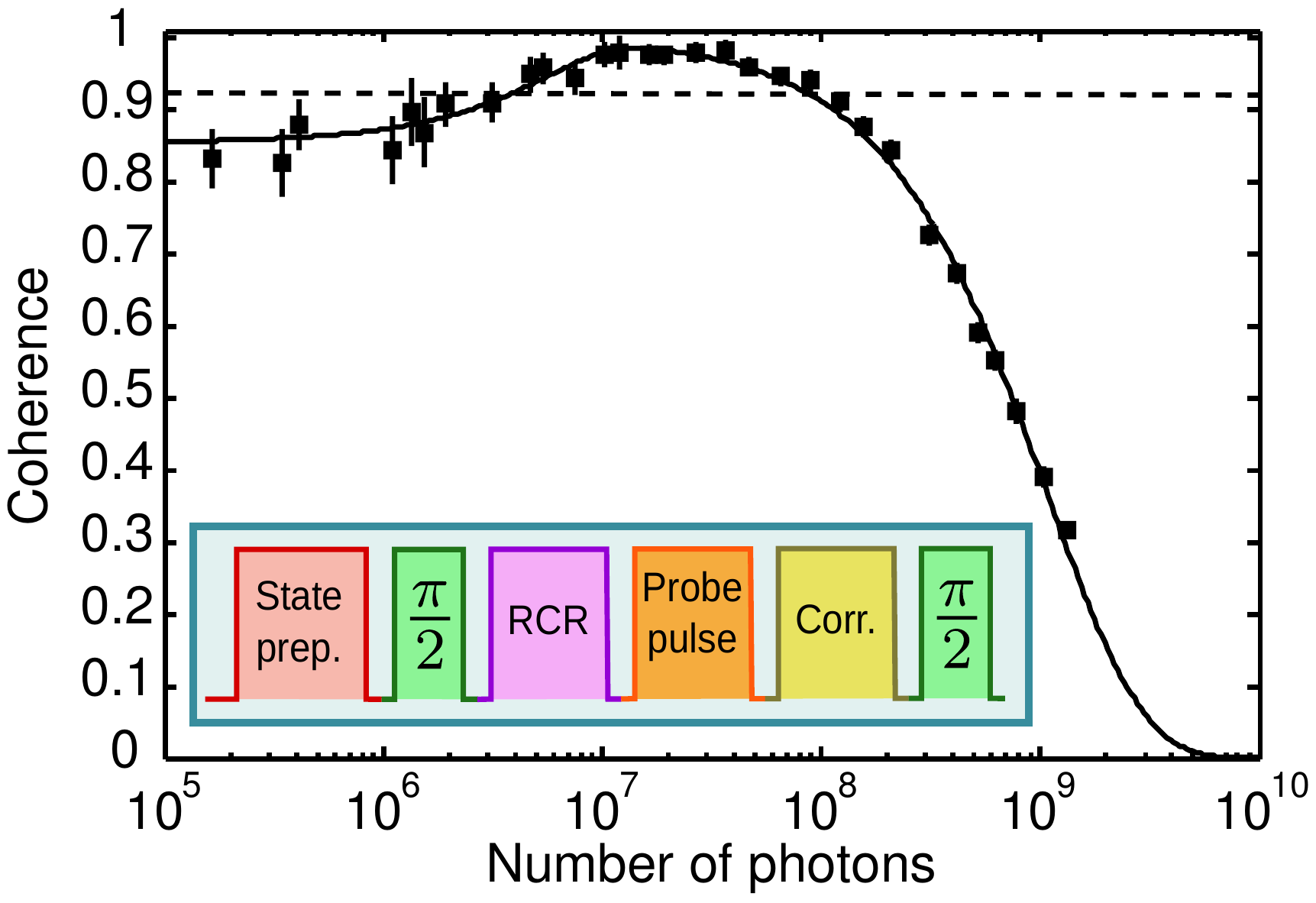}
\caption{(Color online) Coherence recovered after the feedback correction for a RCR angle $\alpha = \pi/8$ versus the number of photons per sideband and per detection pulse. The experimental data result from the average of 50 realizations and the error bars are the $\pm 1 \sigma$ statistical uncertainty. The solid line is a fit with Eq.~(\ref{eq:fit_vs_Np}) and the dashed line is the remaining coherence after the RCR equal to $\cos (\pi/8)$. The inset shows a scheme of the experimental sequence used to determine the coherence thanks to the final $\pi/2$ pulse closing a Ramsey interferometer.}
\label{fig:coherence_vs_Np}
\end{center}
\end{figure}

The output coherence is estimated by adding a $\pi/2$ pulse after the correction pulse, which closes the Ramsey interferometer opened by the initial $\pi/2$ pulse (see inset in Fig.~\ref{fig:coherence_vs_Np}). The value of $J_{z}$ at the interferometer output, averaged over many realizations of the experimental cycle, is an estimate of the remaining coherence. 

The results obtained for a RCR angle $\alpha = \pi/8$ are presented in Fig.~\ref{fig:coherence_vs_Np}. The experimental data are fitted with the function:
\begin{equation}
\eta_{\alpha}^{\rm out} = \left[ p_{s} + \frac{1}{\sqrt{2}} (1-p_{s}) \right] e^{-\gamma N_{p}},
\label{eq:fit_vs_Np}
\end{equation}
where the first factor is the coherence versus the success probability [Eq.~(\ref{eq:p_success})] for $\alpha = \pi/8$ (Tab.~\ref{tab:evol_grand_caract}), and the exponential factor accounts for the spontaneous emission induced by $N_{p}$ photons.

For a low photon number in the probe pulse, the success probability is $p_{s}=1/2$ and the remaining coherence after correction is $\eta_{\pi / 8}^{\rm out} = (1 + 1/\sqrt{2})/2 \sim 0.85$, in good agreement with the experimental observation. An optimal coherence of 0.985 is recovered for $1.4 \times 10^{7}$ photons: this value is higher than the coherence after the RCR ($\eta_{\pi / 8} = \cos \pi/8 \sim 0.924$), which proves the efficiency of the feedback controller according to the coherence criterion defined in Sec.~\ref{sec:feed_eff_crit}.

\begin{figure*}
\includegraphics[width=17cm,keepaspectratio]{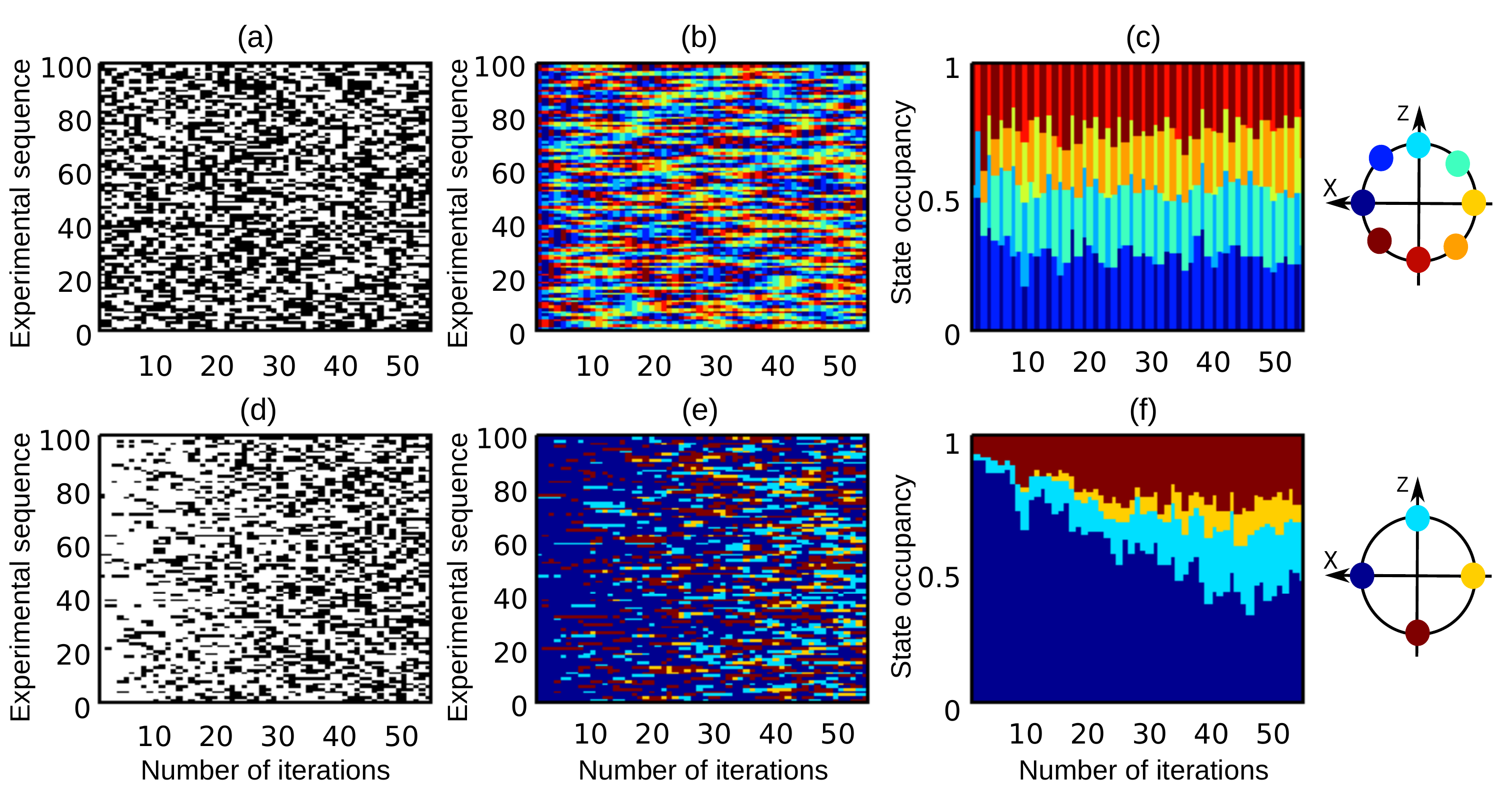}
\caption{(Color online) Data analysis flow for the iterated feedback correction. The RCR angle is $\alpha = \pi/4$, the probe contains $1.4 \times 10^{7}$ photons per sideband, and each iteration lasts 140 $\mu$s. Top: analysis of the state evolution without feedback; (a) sign of the binary RCR set by the random number generator (black $+\alpha$, white $-\alpha$), (b) resulting state, and (c) state occupancy. Bottom: results with feedback correction; (d) controller decisions (right in white, wrong in black), (e) resulting state after the correction, and (f) state occupancy. The legend on the right indicates the pointing direction of state associated with each color.}
\label{fig:analysis_data_seq}
\end{figure*}

\subsection{Iterated feedback correction of a binary RCR}

We now repeat the sequence RCR-correction to study the long-term efficiency of the feedback controller. Characterizing the coherence evolution using the previous method --- adding an extra $\pi/2$ pulse to close a Ramsey interferometer --- would require a large number of experimental cycles, since the remaining coherence must be estimated after each iteration.

However, we previously verified that the coherence reduction on one iteration results from the product of two contributions: one related to the success probability of the RCR detection, and one to the spontaneous emission induced by the probe; the trap induced phase-shift being negligible on a single iteration (see Sec.~\ref{sec:trap_dec}). Therefore, we can estimate the coherence evolution from the independent determination of each process contribution. This method significantly reduces the number of experimental cycle required for the measurement. Here, we present how the data are analyzed to obtain the coherence, and we study the effect of the finite sample size on the estimation.

For each experimental realization, we record both the RCR sign and that of the correction rotation set by the feedback controller: their comparison determines whether the controller took the right decision. The results of many experimental realizations [Fig.~\ref{fig:analysis_data_seq}~(d)] is compared to those obtained in the absence of feedback [Fig.~\ref{fig:analysis_data_seq}~(a)]: during the first 20 iterations the  feedback controller mostly takes the right decision, and it converges towards a success probability of $1/2$ due to the decoherence of the sample when the number of iterations increases. It is important to note that the state can be recovered even after a wrong decision taken by the controller, therefore the feedback scheme is robust against detection and computation errors.

\begin{figure}[!h]
\begin{center}
\includegraphics[width=8.5cm,keepaspectratio]{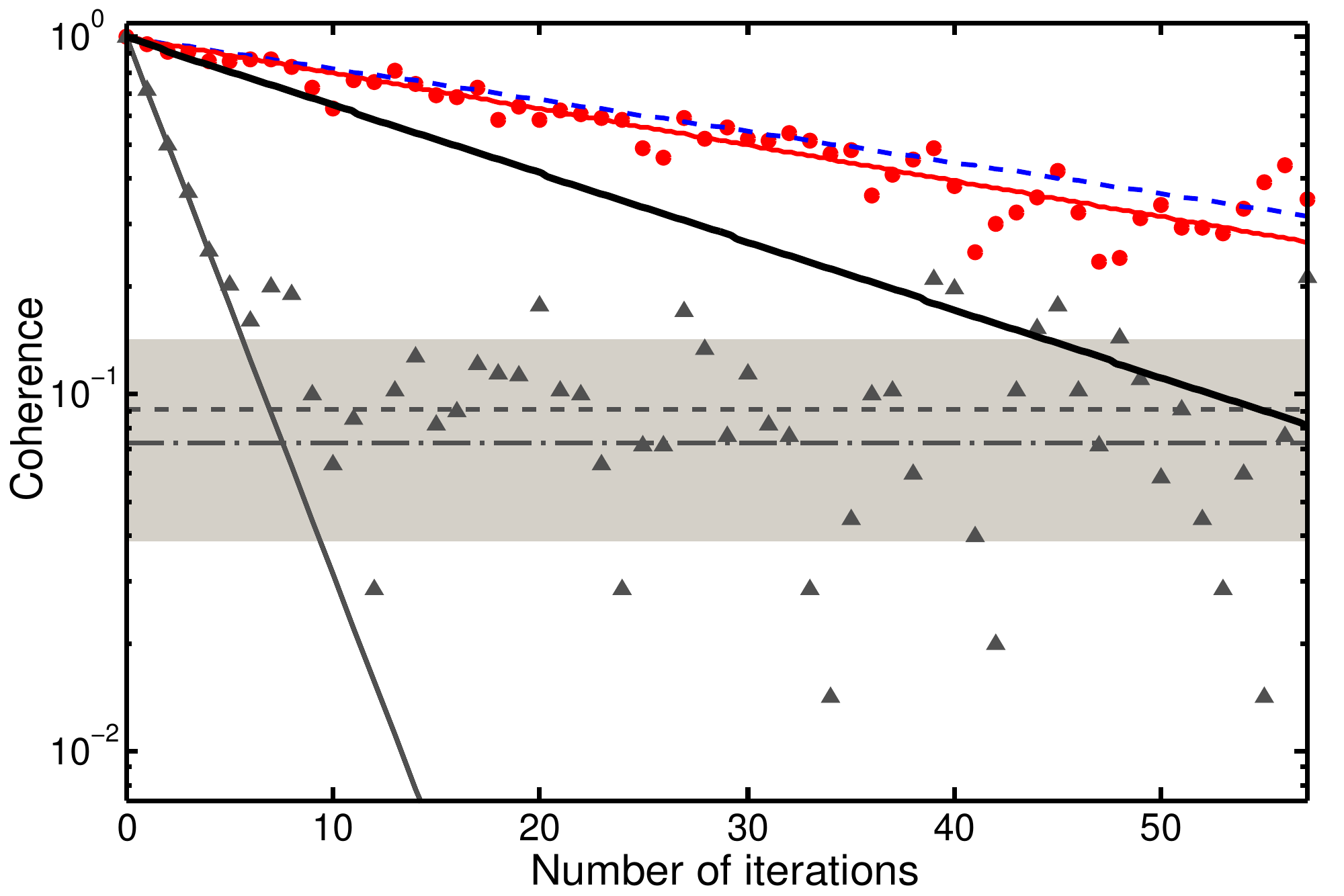}
\caption{(Color online) Evolution of the coherence with the number of iterations, obtained from 100 experimental measurements. The red circles are the coherence resulting from the state occupancy and the solid red line is an exponential fit. The dashed blue line is the coherence loss due to the spontaneous emission induced by the probe. The solid black line is the estimated coherence. The grey triangles are the estimated coherence resulting from the state occupancy in absence of feedback control and the solid grey line is the expected coherence $\eta = 2^{-N_{\rm it}/2}$. The dashed-dotted horizontal grey line is the expected limit on the coherence resolution due to the finite size of the statistical sample [Eq.~(\ref{eq:res_coh_fin_size})]. The dashed horizontal grey line is the average of the experimental data for  $N_{\rm it} > 20$ and the shaded area indicates the related standard deviation.}
\label{fig:coh_vs_nit}
\end{center}
\end{figure}

Since the applied RCRs and corrections are known, we can reconstruct the path followed by the pseudo-spin during the each sequence [Figs.~\ref{fig:analysis_data_seq}~(b) and \ref{fig:analysis_data_seq}~(e)]. This analysis performed over many experimental realizations provides an estimate of the state occupancy $P_{k} (N_{\rm it})$, which is the probability to be in the state $\left| k \right\rangle$ after $N_{\rm it}$ iterations [Figs.~\ref{fig:analysis_data_seq}~(c) and \ref{fig:analysis_data_seq}~(f)]. We see that without feedback the system converges quickly towards equidistributed populations, whereas when feedback is applied the occupancy of the target state dominates for several tens of iterations. The feedback control damps the spin diffusion around the Bloch sphere. The state occupancy in turns allows us to estimate the density operator: $\rho (N_{\rm it}) = \sum_{k} P_{k} (N_{\rm it}) \left| k \right\rangle \left\langle k \right|$. The coherence is finally calculated using Eq.~(\ref{eq:coh_dft}), and shown in Fig.~\ref{fig:coh_vs_nit} (red circles and red line). Note that the spontaneous emission lowers the success probability as the number of detections increases, as shown in Fig.~\ref{fig:analysis_data_seq}~(d), this effect is thus included in the contribution of the state occupancy to the coherence.

The above evaluation of the coherence considers only the contribution resulting from the success probability. The decoherence due to the probe spontaneous emission must be added, using the decoherence rate $\gamma$ (obtained by the analysis performed in Sec.~\ref{sec:coh_vs_nb_photons}) and the total number of photons sent through the atomic sample. The coherence evolution, under the influence of the spontaneous emission only, is depicted in Fig.~\ref{fig:coh_vs_nit} (dashed blue line). Finally, the estimated coherence (Fig.~\ref{fig:coh_vs_nit}, solid black line) is the product of the two previous contributions (state occupancy and spontaneous emission).

We must also consider the contribution of the inhomogeneous differential light-shift induced by the trapping beam between $| 0 \rangle$ and $| 1 \rangle$ (see Sec.~\ref{sec:trap_dec}). This effect is a small correction: it reduces the coherence of about 10~\% since the experiment lasts 7~ms (see Fig.~\ref{fig:contrast_vs_time}). Moreover, it influences in the same way the coherence evolutions with and without feedback, therefore it does not play a role in the understanding of the feedback efficiency.

Fig.~\ref{fig:coh_vs_nit} also displays the evolution of the coherence without feedback (grey triangles) estimated from the state occupancy. We see that, while the first points are following the expected exponential decay $\eta = 2^{-N_{\rm it}/2}$ (solid grey line), a floor is reached for a number of iterations larger than $\sim 10$. This effect can be understood as a result of the  finite number $N$ of experimental realizations, which limits the dynamic range of the coherence measurement.

We now quantify this effect. The probability $P_{k}$ to be in the state $k \in [0,n-1]$ is estimated with an uncertainty whose standard deviation is
\begin{equation}
\delta P_{k} (N) = \sqrt{\frac{P_{k} \left( 1 - P_{k} \right)}{N}}.
\label{eq:deltaP_fin_samp}
\end{equation}
As a consequence, this uncertainty on the probability translates into an uncertainty on the estimated coherence: $\eta = \widetilde{\eta}(N) + \delta \widetilde{\eta}(N)$, where $\eta$ is the actual coherence and $\widetilde{\eta} (N)$ is the value estimated from $N$ samples. 

We consider the situation without feedback corrections, allowing us to analyze this effect on a simple and well understood scenario. As shown in App.~\ref{app:fin_size_samp}, the variance of the estimated coherence satisfies
\begin{equation}
\delta \widetilde{\eta} (N) = \frac{2^{1/4} n^{-3/4} \sqrt{n-1}}{\sqrt{N}}.
\label{eq:res_coh_fin_size}
\end{equation}
Note that, since $\delta \widetilde{\eta} (N) \propto N^{-1/2} \rightarrow 0$ as $N \rightarrow \infty$, $\widetilde{\eta} (N)$ is an estimator of $\eta$. Moreover, it is also interesting to observe that $\delta \widetilde{\eta} \sim n^{-1/2}$ as $n \rightarrow \infty$: this effect is less important when the RCR angle $\alpha$ is smaller. This result is compared with the experimental data in Fig.~\ref{fig:coh_vs_nit}, where the theoretical estimation of $\delta \widetilde{\eta}$ is performed with $n=4$ since at each iteration only four states are populated.

\subsection{\label{corr_analog_RCR}Correction of an analog RCR}

Following Sec.~\ref{sec:th_analog_rcr}, we experimentally demonstrate the active stabilization against an analog RCR, detailing the result stated in Ref.~\cite{vanderbruggen2013}. The sequence consists in an analog RCR uniformly distributed in $[ -\pi/2, \pi/2 ]$, followed by a 1.5~$\mu$s probe pulse containing $2.8 \times 10^{7}$ photons per sideband and a correction microwave pulse. The feedback controller sets both the phase sign and the duration of the correction pulse. We adopt the control strategy where the correction angle is set to be proportional to the $J_{z}$ measurement result, and the gain is optimized for a RCR angle of $\pi/3$. This value is adopted considering the measurement resolution $\sigma = 14$~\% and the analysis performed in Sec.~\ref{sec:corr_strat_2}.

\begin{figure}[h!]
\begin{center}
\includegraphics[width=6cm,keepaspectratio]{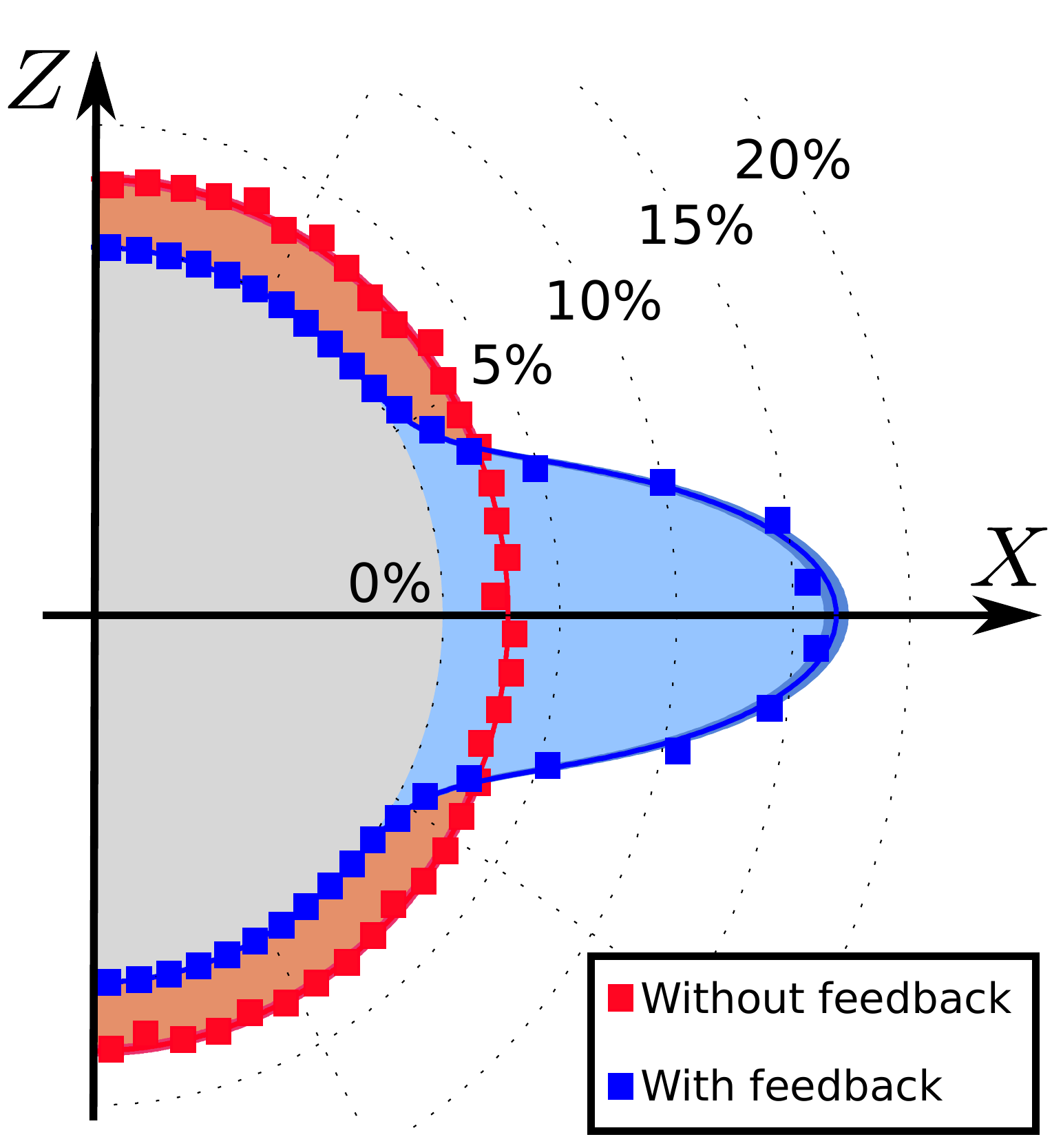}
\caption{(Color online) Probability distribution for the pointing direction of the state in the $(X,Z)$ plane of the Bloch sphere, obtained from 5000 repetitions of the experimental sequence. In red (light gray) is the distribution after an analog RCR uniformly distributed in $[-\pi/2,+\pi/2]$, and in blue (dark gray) is the resulting distribution after feedback correction.}
\label{fig:analog_feed_res}
\end{center}
\end{figure}

From the measurement of the pointing direction of the CSS after the correction versus the direction after the RCR,  we determine the probability distribution of the resulting statistical mixture, shown in Fig.~\ref{fig:analog_feed_res}. We first measured this distribution for an analog RCR uniformly distributed in $[-\pi/2,+\pi/2]$ to verify that the right behavior of the FPGA program coupled to the QNRG. In a second time, the distribution after feedback correction was measured. We see that the distribution is pointing along the $X$ axis, as desired. The spin spread corresponds to a remaining coherence of 0.979. Another 0.979 factor must be consider due to the spontaneous emission from the probe, resulting in a remaining coherence of 0.958 after feedback correction.

\section{Conclusion}

We demonstrated, both theoretically and experimentally, that weak nondestructive measurements can be used to control in real-time the orientation of a CSS. We studied a benchmark scenario based on RCRs and presented its implementation with a collective atomic pseudo-spin. For this purpose, we developed a nondestructive probe, based on FM spectroscopy and using a single beam, able to directly probe the population difference on the clock transition of alkali-metal atoms without inducing any light-shift. We showed that, for an optimal trade-off between spontaneous emission and detection resolution, the feedback control recovers the coherence of the CSS for different kinds of RCRs (binary and analog). Moreover, it can protect over time a CSS subject to repeated RCRs.

The demonstration was performed on a ensemble of trapped alkali-metal atoms. These systems are particularly interesting since trapped ultra-cold atom sensors are at the center of recent developments for long interrogation time embedded interferometers, such as microwave clocks \cite{deutsch2010} or gyroscopes \cite{garrido2012}. The present method is perfectly suited for the real-time monitoring and control of atom interferometers, paving the way towards new interferometric schemes beyond the Ramsey sequence, where interrogation and detection would be mixed. Our feedback control scheme can be used for example to lock the atomic state near the zero phase-shift position, realizing sensors with increased dynamic range and improved stability against large perturbations. In particular, the present feedback control meets the requirements to realize an atomic phase-lock loop \cite{shiga2011}, where not only the frequency but also the phase of an electromagnetic wave would be locked to an atomic reference, dramatically reducing the constraints on the local oscillator phase noise.

\section*{Acknowledgments}

We acknowledge funding from DGA, CNES, the European Union (EU) (iSENSE), EURAMET (QESOCAS), ANR (MINIATOM), and ESF Euroquam. LCFIO and SYRTE are members of the Institut Francilien de Recherche sur les Atomes Froids (IFRAF). P.~B. acknowledges support from a chair of excellence of R{\'e}gion Aquitaine. E.~C. acknowledges support from Quantel.

\appendix

\section{\label{app:crit_param}Derivation of the efficiency parameters}

In this appendix, we calculate the parameters used to quantify the feedback efficiency: coherence, von Neumann entropy and  fidelity.

\subsection{Coherence}

\subsubsection{Coherence of a pure CSS}

The mean Bloch vector related to a CSS $\left| \theta, \varphi \right\rangle$ containing $j = N_{\rm at}/2$ atoms is the vector pointing in the $\left( \theta, \varphi \right)$ direction:
\begin{equation}
\left\langle \mathbf{J} \right\rangle_{\theta,\varphi} = \left(
\begin{array}{c}
\left\langle J_{x} \right\rangle_{\theta,\varphi} \\
\left\langle J_{y} \right\rangle_{\theta,\varphi}   \\
\left\langle J_{z} \right\rangle_{\theta,\varphi}
\end{array}
\right) = j \left(
\begin{array}{c}
- \sin \theta \cos \varphi \\
\sin \theta \sin \varphi   \\
- \cos \theta
\end{array}
\right).
\end{equation}
The coherence of a CSS is thus $\left\| \left\langle \mathbf{J} \right\rangle_{\theta, \varphi} \right\|/j = 1$, which proves that the coherence of the initial state $\left| \psi_{0} \right\rangle = \left| \pi/2,0 \right\rangle$ is unitary.

\subsubsection{\label{app:coh_mixt_css} Coherence of a statistical mixture of CSSs}

Let $\left\{ \left| \theta_{k}, \varphi_{k} \right\rangle \right\}$ be a set of CSSs, an arbitrary statistical mixture of these states is described by a density matrix of the form $\rho = \sum_{k} p_{k} \left| \theta_{k}, \varphi_{k} \right\rangle \left\langle \theta_{k}, \varphi_{k} \right|$, where $\sum_{k} p_{k} = 1$. By the linearity of the trace, we have for $l=x,y,z$:
\begin{eqnarray}
\left\langle J_{l} \right\rangle (\rho) & = & \mathrm{Tr} \left( J_{l} \rho \right) \\
& = & \sum_{k} p_{k} \mathrm{Tr} \left( J_{l} \left| \theta_{k}, \varphi_{k} \right\rangle \left\langle \theta_{k}, \varphi_{k} \right| \right) \\
& = & \sum_{k} p_{k} \left\langle J_{l} \right\rangle_{ \theta_{k}, \varphi_{k} },
\end{eqnarray}
therefore, the mean Bloch vector related to the density operator $\rho$ is $\left\langle \mathbf{J} \right\rangle (\rho) = \sum_{k} p_{k} \left\langle \mathbf{J} \right\rangle_{\theta_{k}, \varphi_{k}}$. As a consequence, the coherence of the mixture $\rho$, $\eta (\rho) = \left\| \left\langle \mathbf{J} \right\rangle (\rho) \right\|/j$, takes the explicit form:
\begin{eqnarray}
\eta (\rho) & = & \left[ \left( \textstyle\sum_{k} p_{k} \sin \theta_{k} \cos \varphi_{k} \right)^{2} + \left( \textstyle\sum_{k} p_{k} \sin \theta_{k} \sin \varphi_{k} \right)^{2} \right. \nonumber \\
& & \left.  + \left( \textstyle\sum_{k} p_{k} \cos \theta_{k} \right)^{2} \right]^{1/2}.
\label{eq:coh_gen}
\end{eqnarray}
Using this relation and the expressions of the density matrix $\mathcal{E}_{\alpha} \left( \rho_{0} \right)$ (Eq.~(\ref{eq:rho_bit_flip})) and $\mathcal{C}_{\alpha} \left( \rho_{0} \right)$ (Eq.~(\ref{eq:rho_out})), it is straightforward to obtain the values of the coherence given in Tab.~\ref{tab:evol_grand_caract}.

Note that in the case $\varphi_{k} = 0$, one obtain the simple relation:
\begin{equation}
\eta (\rho) = \left| \sum_{k} p_{k} e^{i \theta_{k}} \right|.
\label{eq:coh_dft}
\end{equation}
In particular, if $\theta_{k} = 2 \pi k/n$ then the coherence is the modulus of the discrete Fourier transform of the probability distribution $p_{k}$.

All these results obtained for a discrete probability distribution $p_{k}$ can be generalized without difficulty to a continuous distribution $p (\theta)$ ($-\pi \leq \theta \leq \pi$) by the replacement $\sum_{k} \rightarrow \int d\theta$.

\subsection{Fidelity}

The fidelity with respect to an initial state $ \left| \psi_{0} \right\rangle = \left| \theta_{0},\varphi_{0} \right\rangle$ is:
\begin{equation}
\mathcal{F} \left( \rho, \left| \psi_{0} \right\rangle \right) = \left\langle \psi_{0} \right| \rho \left| \psi_{0} \right\rangle = \sum_{k} p_{k} \; \left| \left\langle \theta_{0},\varphi_{0} \right| \left. \theta_{k},\varphi_{k} \right\rangle \right|^{2}.
\label{eq:fidelity_app}
\end{equation}
Expending the CSS $\left| \theta, \varphi \right\rangle$ in the Dicke basis $\left| j,m \right\rangle$ provides \cite{arecchi72}:
\begin{equation}
\left| \theta, \varphi \right\rangle = \sum_{m=-j}^{j} \binom{2j}{j+m}^{\frac{1}{2}} \!\! \sin^{j+m}\frac{\theta}{2} \cos^{j-m}\frac{\theta}{2} e^{-i \varphi m} \left| j,m \right\rangle,
\end{equation}
We then obtain the overlap between two CSSs:
\begin{equation}
\left| \left\langle \theta, \varphi \right| \left. \theta', \varphi' \right\rangle \right| = \left| \cos \frac{\theta}{2} \cos \frac{\theta'}{2} + e^{i \left( \varphi - \varphi' \right)} \sin \frac{\theta}{2} \sin \frac{\theta'}{2} \right|^{2j},
\end{equation}
which can be approximated with a Gaussian distribution, for a large number of atoms ($j\gg1$):
\begin{equation}
\left| \left\langle \theta, \varphi \right| \left. \theta', \varphi' \right\rangle \right|^{2} \sim e^{-\frac{j}{2} \left[ \left( \theta-\theta' \right)^{2} + \frac{1}{2} \left( 1-\cos 2 \theta' \right) \left( \varphi - \varphi' \right)^{2} \right]}.
\label{eq:Q_func_css}
\end{equation}
Finally, using Eqs.~(\ref{eq:fidelity_app}) and (\ref{eq:Q_func_css}), one can evaluate the fidelity of the mixture $\rho$.

\subsection{Von Neumann entropy}

From Eq.~(\ref{eq:Q_func_css}), it appears that if the RCR angle is large enough so that the angles between the CSSs $\left| \theta_{k}, \varphi_{k} \right\rangle$ in the statistical mixture are large compared to the atomic shot-noise, that is, if they satisfy:
\begin{equation}
\forall k \neq k', \; \left\{
\begin{array}{c}
\theta_{k} - \theta_{k'} \gg 1/\sqrt{N_{\rm at}} \\
\mathrm{or} \\
\varphi_{k} - \varphi_{k'} \gg 1/\sqrt{N_{\rm at}}
\end{array}
\right.
\end{equation}
then the states are almost orthogonal $\left\langle \theta_{k}, \varphi_{k} \right| \left. \theta_{k'}, \varphi_{k'} \right\rangle \sim \delta_{k,k'}$, and the von Neumann entropy of $\rho$ satisfies:
\begin{equation}
S (\rho) \simeq - \sum_{k} p_{k} \log_{2} p_{k},
\end{equation}
in other words, since here the RCR angles are larger than the atomic shot-noise, only the pointing direction of the CSS matters and the von Neumann entropy results to be the Shannon entropy.

Once again, combining this relation and the expressions of the density matrix $\mathcal{E}_{\alpha} \left( \rho_{0} \right)$ [Eq.~(\ref{eq:rho_bit_flip})] and $\mathcal{C}_{\alpha} \left( \rho_{0} \right)$ [Eq.~(\ref{eq:rho_out})], provides the values of the von Neumann entropy in Tab.~\ref{tab:evol_grand_caract}.

\section{\label{sec:trap_light_shift_clock}Trap shift on the clock transition}

We estimate the differential light-shift induced by the trap radiation between the two hyperfine sub-levels of the 5$^{2}$S$_{1/2}$ state. The trapping radiation couples mainly the fundamental levels to the excited states 5$^{2}$P$_{1/2}$ and 5$^{2}$P$_{3/2}$. The transitions and the notations used in the calculation are introduced in Fig.~\ref{fig:contrast_vs_time}.

The light-shift induced on the hyperfine state $\left| 5^{2}\mathrm{S}_{1/2}, F \right\rangle$ is \cite{grimm2000}:
\begin{align}
\frac{\Delta_{F} \left( \mathbf{r} \right)}{I\left( \mathbf{r} \right)} &= \frac{\pi c^{2}}{2} \left[ \frac{ S_{1/2}^{1/2} \, \Gamma_{1/2}}{\omega_{F,\frac{1}{2}}^{3} \left( \omega - \omega_{F,\frac{1}{2}} \right)} + \frac{S_{1/2}^{3/2} \, \Gamma_{3/2}}{\omega_{F,\frac{3}{2}}^{3} \left( \omega - \omega_{F,\frac{3}{2}} \right)} \right],
\label{eq:light_shift_diff_fund_tot}
\end{align}
where $\omega$ is the frequency of the trap radiation, $S_{J}^{J'} = (2J'+1)/(2J+1)$ and $I\left( \mathbf{r} \right)$ is the intensity profile of the trap beam.

The hyperfine splitting between the $F=1$ and $F=2$ levels is $\Delta_{\rm HF} = \omega_{2,1/2}-\omega_{1,1/2} = \omega_{2,3/2}-\omega_{1,3/2}$, and since the following hypothesis are satisfied for $J'=1/2, \; 3/2$ : $\Delta_{\rm HF} \ll \omega_{1,J'}, \; \omega_{2,J'}, \; \omega-\omega_{1,J'}$, Eq.~(\ref{eq:light_shift_diff_fund_tot}) can be expanded to the first order in $\Delta_{\rm HF}$. The differential shift, $\delta \left( \mathbf{r} \right) = \Delta_{2} \left( \mathbf{r} \right) - \Delta_{1} \left( \mathbf{r} \right)$, can thus be written as
\begin{align}
\delta \left( \mathbf{r} \right) &= \frac{\pi c^{2}}{2} \Delta_{\rm HF} \left[ \frac{\Gamma_{1/2}}{\omega_{1/2}^{3} \delta_{1/2}} \left( \frac{1}{\delta_{1/2}} - \frac{3}{\omega_{1/2}} \right) \right. \nonumber \\
& \;\;\;\;\;\;\;\;\;\;\; \left. + \frac{2 \Gamma_{3/2}}{\omega_{3/2}^{3} \delta_{3/2}} \left( \frac{1}{\delta_{3/2}} - \frac{3}{\omega_{3/2}} \right) \right] I \left( \mathbf{r} \right),
\end{align}
where $\omega_{J'} \equiv \omega_{1,J'}$ and $\delta_{J'} \equiv \omega - \omega_{J'}$. 

The wavelengths of the relevant transitions are $\{ \lambda_{1/2}, \lambda_{3/2} \} = \{ 795, 780 \}$~nm, and the related linewidths are $\{ \Gamma_{1/2}, \Gamma_{3/2} \} = 2 \pi \times \{ 5.746, 6.065\}$~MHz, moreover the hyperfine splitting is $\Delta_{\rm HF} = 2 \pi \times 6.834$~GHz \cite{steck}. Therefore, the differential shift at the trap center for an optical power of 10~W per cavity arm and a waist of 100~$\mu$m at 1550~nm is $\delta \sim 54$~Hz.


\section{\label{app:fin_size_samp}Derivation of the estimated coherence uncertainty}

Assuming independent processes, the variance of the estimated coherence is:
\begin{equation}
\delta \widetilde{\eta}^{2} (N) = \sum_{k=0}^{n-1} \left( \frac{\partial \eta}{\partial P_{k}} \right)^{2} \delta P_{k}^{2} (N).
\label{eq:var_coh_fin_samp}
\end{equation}
Using Eq.~(\ref{eq:coh_dft}) relating $\eta$ to $P_{k}$, we obtain:
\begin{equation}
\eta \frac{\partial \eta}{\partial P_{k}} = \left( \sum_{i=0}^{n-1} P_{i} \cos \theta_{i} \right) \cos \theta_{k} + \left( \sum_{i=0}^{n-1} P_{i} \sin \theta_{i} \right) \sin \theta_{k},
\label{eq:inter_coh_fin_samp}
\end{equation}
where $\theta_{k} = 2 \pi k/n$ and $n=2 \pi / \alpha$, for a RCR angle $\alpha$. 

We consider the coherence evolution without feedback correction: in this case the coherence drops rapidly with the number $N_{\rm it}$ of iterations, since $\eta = (\cos \alpha)^{N_{\rm it}} = 2^{-N_{\rm it}/2}$ for $\alpha=\pi /4$. Thus for $N_{\rm it}$ sufficiently large, the estimated coherence is limited by the uncertainty in the estimation of the probabilities: $\delta \widetilde{\eta} (N) \gg \widetilde{\eta} (N)$. Moreover, since no feedback is applied, the state convergences rapidly towards an uniformly distributed mixture of the states $| k \rangle$: $P_{k} = 1/n$. In these conditions, injecting Eqs.~(\ref{eq:inter_coh_fin_samp}) and (\ref{eq:deltaP_fin_samp}) into Eq.~(\ref{eq:var_coh_fin_samp}) provides Eq.~(\ref{eq:res_coh_fin_size}).

\end{document}